\documentclass[acmtog,authorversion]{acmart}

\usepackage[utf8]{inputenc}

\AtBeginDocument{%
  \providecommand\BibTeX{{%
    \normalfont B\kern-0.5em{\scshape i\kern-0.25em b}\kern-0.8em\TeX}}}

\setcopyright{acmcopyright}\acmJournal{TOG}
\acmYear{2020}\acmVolume{39}\acmNumber{6}\acmArticle{208}\acmMonth{12} \acmDOI{10.1145/3414685.3417843}

\ccsdesc[500]{Applied computing~Computer-aided design}
\ccsdesc[500]{Computing methodologies~Shape modeling}
\ccsdesc[500]{Computing methodologies~Mixture modeling}

\acmSubmissionID{338} 

\usepackage[english]{babel}
\usepackage{csquotes}
\usepackage{subcaption}
\usepackage{color}
\usepackage{hyperref}
\usepackage[inline]{enumitem}
\usepackage{amsfonts}
\usepackage{amsmath}
\usepackage{commath}
\usepackage{overpic}
\usepackage{contour}
\usepackage{soul}
\usepackage[super]{nth}

\usepackage{bm}

\newcommand{\RR}{\mathbb{R} }

\def\stressbar{\begin{overpic}[width=1mm,angle=-90]{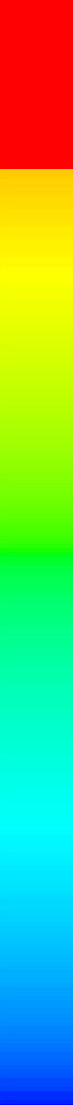}
	\put(0,12){\scriptsize$0$}
	\put(35,12){\scriptsize MPa}
	\put(85,12){\scriptsize$>65$}
	\end{overpic}}
	
\def\kinkbar{\begin{overpic}[width=1mm,angle=-90]{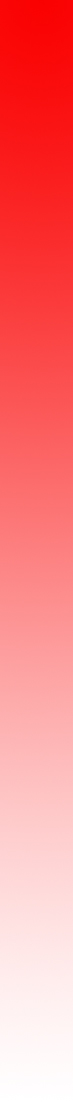}
	\put(0,12){\scriptsize$0^\circ$}
	\put(35,12){\scriptsize kink}
	\put(83,12){\scriptsize$>35^\circ$}
	\end{overpic}}

\def\offsetbar{\begin{overpic}[width=1mm,angle=-90]{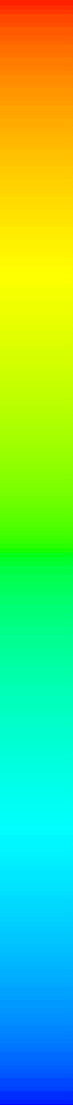}
	\put(0,12){\scriptsize$0$}
	\put(35,12){\scriptsize mm}
	\put(83,12){\scriptsize$0.12$}
	\end{overpic}}

\DeclareMathOperator{\Tr}{Tr}

\begin{document}

%%%%% TITTLE %%%%%

\title{Computational Design of Cold Bent Glass Façades}

\author{Konstantinos Gavriil}
\authornote{Joint first authors}
\affiliation{\institution{TU Wien}}

\author{Ruslan Guseinov}
\authornotemark[1]
\affiliation{\institution{IST Austria}}
\email{guseynov.ruslan@gmail.com}

\author{Jes{\'u}s P{\'e}rez}
\affiliation{\institution{URJC}}

\author{Davide Pellis}
\affiliation{\institution{TU Wien}}

\author{Paul Henderson}
\affiliation{\institution{IST Austria}}
\email{https://www.pmh47.net}

\author{Florian Rist}
\affiliation{\institution{KAUST}}

\author{Helmut Pottmann}
\affiliation{\institution{KAUST, TU Wien}}
\email{helmut.pottmann@kaust.edu.sa}

\author{Bernd Bickel}
\affiliation{\institution{IST Austria}}
\email{bernd.bickel@ist.ac.at}

% \affiliation{%

\begin{abstract}
Cold bent glass is a promising and cost-efficient method for realizing doubly curved glass façades. They are produced by attaching planar glass sheets to curved frames and must keep the occurring stress within safe limits. However, it is very challenging to navigate the design space of cold bent glass panels because of the fragility of the material, which impedes the form finding for practically feasible and aesthetically pleasing cold bent glass façades. We propose an interactive, data-driven approach for designing cold bent glass façades that can be seamlessly integrated into a typical architectural design pipeline. Our method allows non-expert users to interactively edit a parametric surface while providing real-time feedback on the deformed shape and maximum stress of cold bent glass panels. The designs are automatically refined to minimize several fairness criteria, while maximal stresses are kept within glass limits. We achieve interactive frame rates by using a differentiable Mixture Density Network trained from more than a million simulations. Given a curved boundary, our regression model is capable of handling multistable configurations and accurately predicting the equilibrium shape of the panel and its corresponding maximal stress. We show that the predictions are highly accurate and validate our results with a physical realization of a cold bent glass surface.
\end{abstract}

\keywords{mechanical simulation, cold bent glass, neural networks, computational design, inverse design}

\begin{teaserfigure}
  \includegraphics[width=\textwidth]{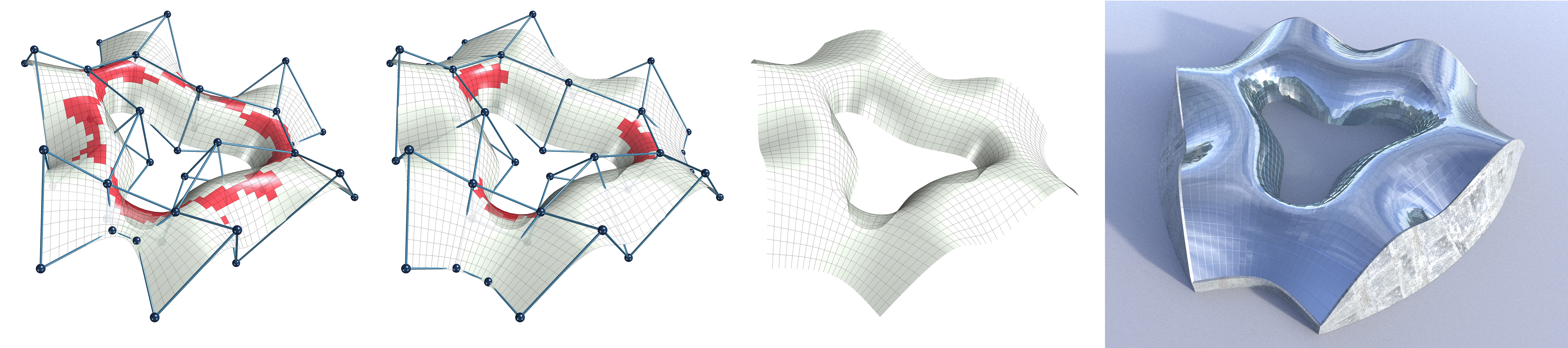} %the other one is teaser_03.jpg
  \caption{Material-aware form finding of a cold bent glass façade. From left to right: initial and revised panel layouts from an interactive design session with immediate feedback on the glass shape and maximum stress (red color indicates panel failure). The surface design is then optimized for stress reduction and smoothness. The final façade realization using cold bent glass features doubly curved areas and smooth reflections. }
  \label{fig:teaser}
\end{teaserfigure}

\maketitle

%%%%% BODY %%%%%

%color scheme https://coolors.co/e63946-f1faee-a8dadc-457b9d-1d3557

\section{Introduction}

\begin{figure}[t!]
    \centering
    \includegraphics[trim=15 13 0 10,clip, width=0.3\columnwidth]{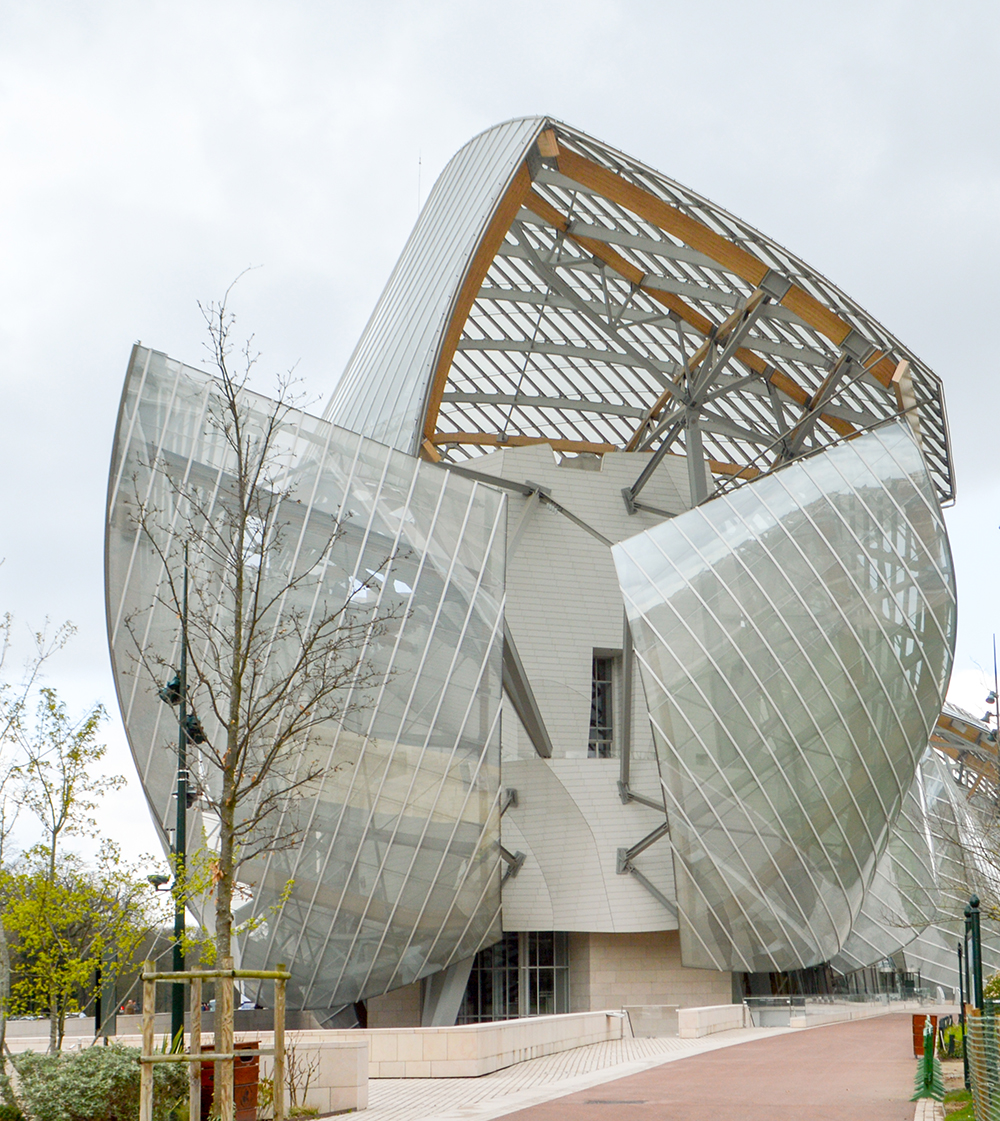}
    \includegraphics[trim=180 0 230 0,clip, width=0.359\columnwidth]{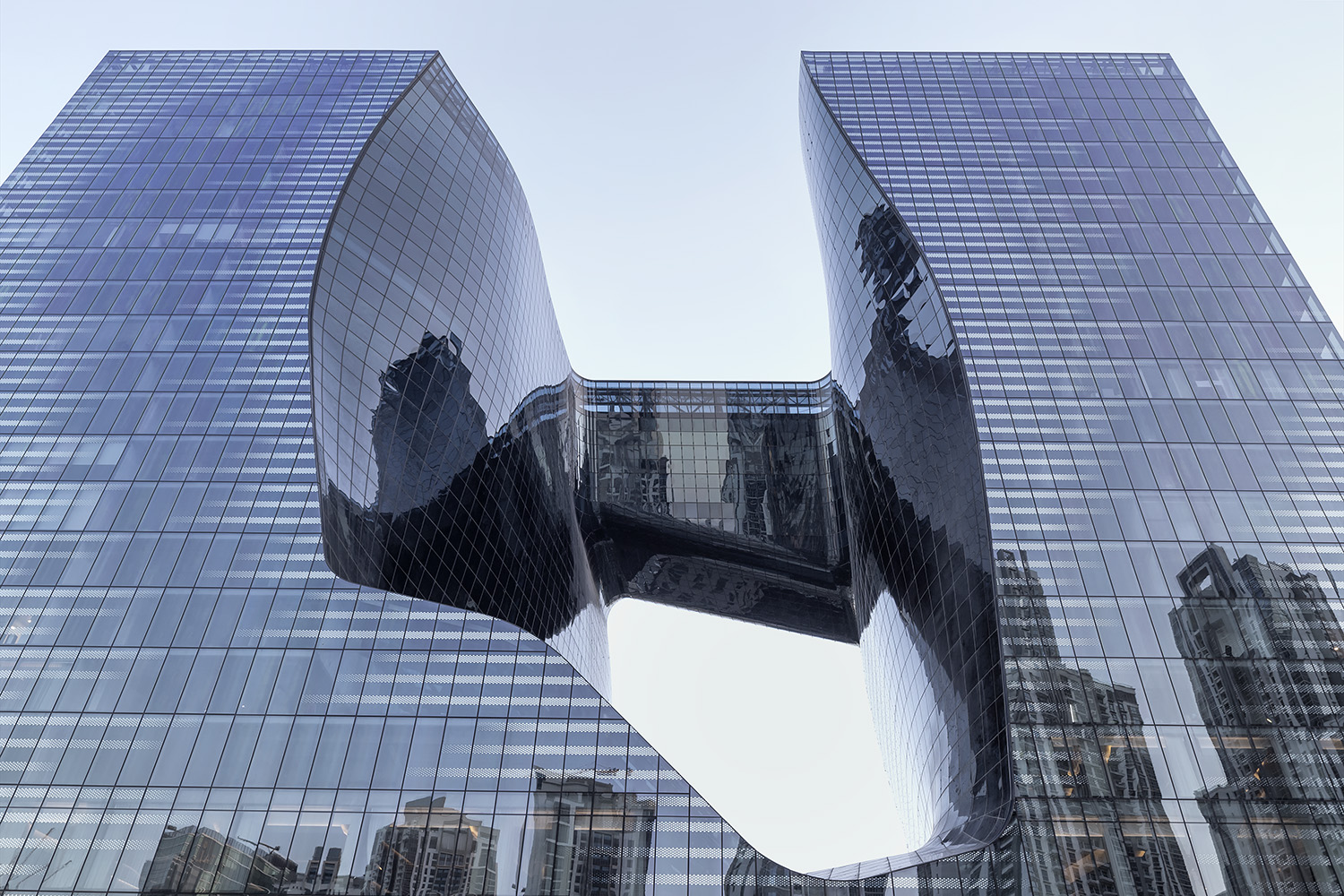}
    \begin{minipage}[b]{.32\linewidth}
\caption{Examples of curved glass façades. Left: Fondation Louis Vuitton, Paris, by Frank Gehry. Right: Opus, Dubai, by Zaha Hadid Architects (Photo: Danica O. Kus).}
\label{fig:glassbuildings}
\end{minipage}
\end{figure}

Curved glass façades allow the realization of aesthetically stunning looks for architectural masterpieces, as shown in Figure~\ref{fig:glassbuildings}. The curved glass is usually made with hot bending, a process where the glass is heated and then formed into a shape using a mold or using tailored bending machines for spherical or cylindrical shapes. While being able to unleash these stunning designs from being restricted to flat panels, this process is laborious and expensive and, thus, an economic obstacle for the realization of exciting concepts such as the NHHQ skyscraper project by Zaha Hadid Architects (Figure~\ref{fig:nhhq_stress}). As a cost-effective alternative, in recent years, architects have started exploring cold bending~\cite{beer2015structural}. Here, planar glass sheets are deformed by mechanically attaching them to a curved frame. Cold bending introduces a controlled amount of strain and associated stress in the flat glass at ambient temperatures to create doubly curved shapes~\cite{datsiou2017design}. Compared with hot bent glass, it has the advantage of higher optical and geometric quality, a wide range of possibilities regarding printing and layering, the usage of partly tempered or toughened safety glass, and the possibility of accurately estimating the stresses from deformation~\cite{belis2007cold,fildhuth2011geometrie}. Furthermore, it reduces energy consumption and deployment time because no mold, heating of the glass, nor elaborate transportation are required.

However, designing cold bent glass façades comes with a challenging form-finding process. How can we identify a visually pleasing surface that meets aesthetic requirements such as smoothness between panels while ensuring that the solution is physically feasible and manufacturable? Significant force loads can occur at the connection between the glass and frame, and it is essential that the deformation of the glass stays within safe limits to prevent it from breaking.

We propose an interactive, data-driven approach for designing cold bent glass façades. Starting with an initial quadrangulation of a surface, our system provides a supporting frame and interactive predictions of the shape and maximum stress of the glass panels. Following a designer-in-the-loop optimization approach, our system enables users to quickly explore and automatically optimize designs based on the desired trade-offs between smoothness, maximal stress, and closeness to a given input surface. Our workflow allows users to work on the 3D surface and the frame only, liberating the designer from the need to consider or manipulate the shape of flat panels – the optimal shape of the flat rest configuration of the glass panels is computed automatically.

At a technical level, we aim to determine the minimum energy states of glass panels conforming to the desired boundary without knowing their rest configuration. Based on extensive simulations of more than a million panel configurations with boundary curves relevant for our application domain, we observed the existence of several (in most cases up to two) stable states for many boundary curves. Identifying both {\it minimum energy states without knowing the rest configuration} and potentially {\it multiple stable states} is a non-trivial problem and cannot be easily computed using standard simulation packages. Furthermore, as a prerequisite for enabling {\it interactive design} for glass façades, we need to solve this problem for hundreds of panels within seconds. 

To achieve these goals, we have developed a learning-based method utilizing a deep neural network architecture and Gaussian mixture model that accurately predicts the shape and maximum stress of a glass panel given its boundary. The training data for the network is acquired from a physics-based shape optimization routine. The predictions of the trained network not observed originally are re-simulated and used for database enrichment. Our model is differentiable, fast enough to interactively optimize and explore the shape of glass façades consisting of hundreds of tiles, and tailored to be easily integrated into the design workflow of architects. As a proof of concept, we have integrated our system into Rhino. We have carefully validated the accuracy and performance of our model by comparing it to a real-world example, and demonstrate its applicability by designing and optimizing multiple intricate cold bent glass façades.

\begin{figure*}[t!]
    \centering
   \includegraphics[trim=0 0 0 0, clip, width=\linewidth]{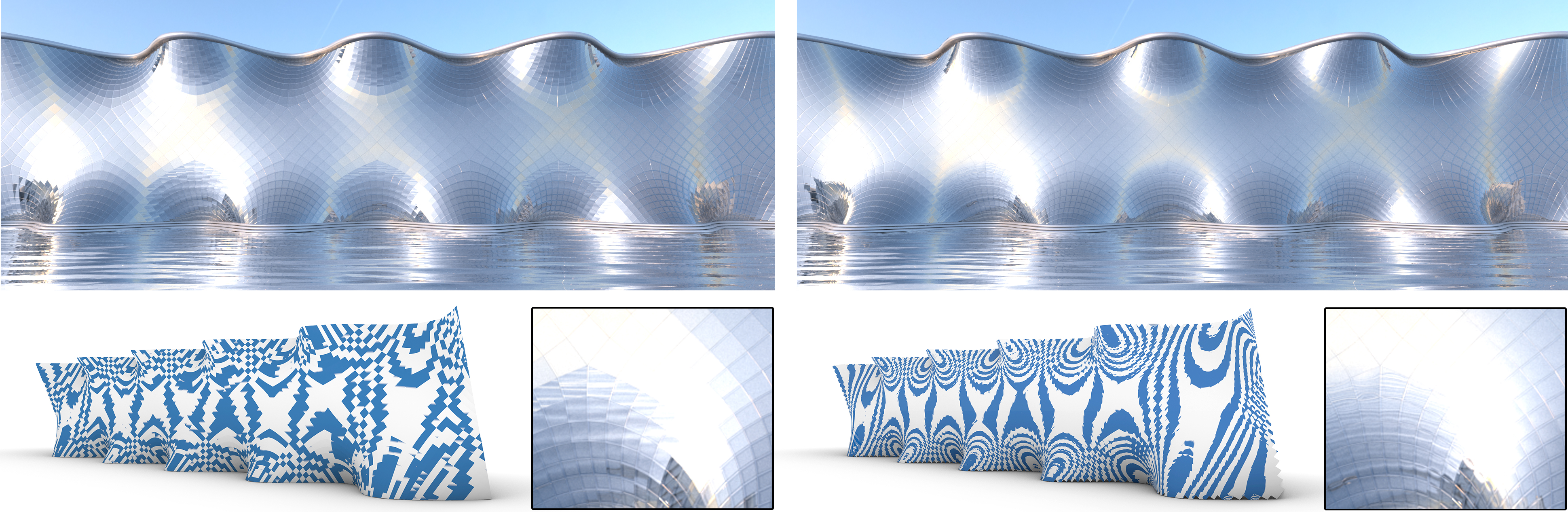}
    \caption{Doubly curved surface panelized using a planar quad mesh following the principal curvature network (left). This is the smoothest possible panelization of this surface achievable with flat panels \cite{pellis-smooth-2019}. The solution using cold bent glass panels designed with our method (right) shows much smoother results. The bottom pictures show the corresponding zebra stripping for both solutions; clearly, smoother stripes are indicators of  higher visual smoothness.}
    \label{fig:curtain_full_comp}
\end{figure*}

\section{Related Work}

Interactive design and shape optimization are areas that have a considerable history in Engineering~\cite{christensen2008introduction}, architecture~\cite{adriaenssens2014shell}, and computer graphics research~\cite{bermano2017state,bickel2018state}, including tools for designing a wide variety of physical artifacts, such as furniture~\cite{umetani2012furniture}, cloth~\cite{Wolff:Aligned-Seams:2019}, robotics~\cite{megaro2015interactive}, and structures for architecture~\cite{eigensatz2010paneling}.
 
Motivated by the digitalization of manufacturing, there is an increased need of computational tools that can predict and support optimizing the physical performance of an artifact during the design process. Several approaches have been developed to guarantee or improve the structural strength of structures~\cite{stava2012stress,ulu2017lightweight}. Focusing on shell-like structures, Musialski et al.~\shortcite{Musialski2015} optimized their thickness such that it minimizes a provided objective function. More recently, Zhao et al.~\shortcite{zhao2017stress} proposed a stress-constrained thickness optimization for shells, and Gilureta et al.~\shortcite{gilureta2019structurally} computed a rib-like structure for reinforcing shells, that is, adding material to the shell to increase its resilience to external loads. Considering both aesthetic and structural goals, Schumacher et al.~\shortcite{schumacher2016stenciling} designed shells with an optimal distribution of artistic cutouts to produce a stable final result. Although we share the general goal of structural soundness, in our problem setting, we cannot change the thickness or material distribution. Additionally, even just determining the feasibility of a desired bent glass shape requires not only solving a forward simulation problem, but also an inverse problem because the rest shape of the glass panel is a priori unknown. Finding an optimal rest shape is often extremely important. Schumacher et al.~\shortcite{schumacher2018set} investigated sandstone as building material that is weak in tension, thus requiring computing an undeformed configuration for which the overall stress is minimized. Similarly, glass panels have a low tensile strength and are subject to very high compression loads during the assembly process, which motivates the need for identifying minimal energy panels.

Notably, several methods have recently been proposed to design doubly curved objects from flat configurations ~\cite{dupeloux2013,guseinov2017curveups,malomo2018flexmaps,konakovic2018rapid,panetta2019x}. However, all these methods rely on significantly more elastic materials and are not targeted for use within an interactive design pipeline. In our application, having an accurate estimation of the stress is critical to predict panel failure and interactively guide designers towards feasible solutions. The need to bridge the gap between accuracy and efficiency motivates the use of a data-driven approach.

\paragraph{Computational design of façades}

Covering general freeform surfaces with planar quadrilateral panels is a fundamental problem in architectural geometry and has received much attention~\cite{glymph2004parametric,liu+2006,liu:conjugate,mesnil2017marionette,pottmann-2015-ag}. The difficulties lie in the close relationship between the curvature behavior of the reference surface and the possible panel layouts. Problems occur especially in areas of negative curvature and if the design choices on the façade boundaries are not aligned with the curvature constraints imposed by planar quad meshes (Figure~\ref{fig:curtain_full_comp}). Using triangular panels, the problems are shifted toward the high geometric complexity of the nodes in the support structure \cite{pottmann-2015-ag}. Eigensatz et al.~\shortcite{eigensatz2010paneling} formulated relevant aspects for architectural surface paneling into a minimization problem that also accounts for re-using molds, thereby reducing production costs. Restricting the design to simple curved panels, Pottmann et al.~\shortcite{pottmann2008freeform} presented an optimization framework for covering freeform surfaces by single-curved (developable) panels arranged along surface strips. However, glass does not easily bend into general developable shapes, limiting the applicability of this technique for paneling with glass. 

A recent alternative for manufacturing doubly curved panels is cold bending of glass. A detailed classification and description of the performance of cold bent glass can be found in~\cite{datsiou2017design}. Eversmann et al.~\shortcite{eversmann2016CompModeling} explored simulations based on a particle-spring method and a commercially available FE analysis tool. Furthermore, they compared the resulting geometries to the measurements of the physical prototypes. For designing multi-panel façade layouts, Eversmann et al.~\shortcite{eversmann2016StructuralPotential} calculated the maximum Gaussian curvature for a few special types of doubly curved panels. This defined a minimal bending radius for exploring multi-panel façade layouts. Berk and Giles~\shortcite{BERK201736} developed a method for freeform surface approximation using quadrilateral cold bent glass panels. However, they limited their fabricability studies to two modes of deformation. Although conceptually simple, we found these approaches too limiting for general curved panels and, thus, have based our approach on a data-driven method. 

\paragraph{Machine learning for data-driven design}

Finite element methods (FEM) are widely used in science and engineering for computing accurate and realistic results. Unfortunately, they are often slow and, therefore, prohibitive for real‐time applications, especially in the presence of complex material behavior or detailed models. 

Dimensionality reduction is a powerful technique for improving simulation speed. Reduced space methods, for example, based on modal analysis~\cite{pentland1989good,barbivc2005real}, are often used to construct linear subspaces, assuming that the deformed shape is a linear combination of precomputed modes. Simulations can then be performed in the spanned subspace, which, however, limits its accuracy, especially in the presence of non-linear behavior. Non-linear techniques such as numerical coarsening~\cite{chen2015data} allow for the reduction of the models with inhomogeneous materials, but usually require precomputing and adjusting the material parameters or shape functions~\cite{chen2018numerical} of the coarsened elements. Recently, Fulton et al.~\shortcite{fulton2019latent} proposed employing autoencoder neural networks for learning nonlinear reduced spaces representing deformation dynamics. Using a full but linear simulation, NNWarp ~\cite{luo2018nnwarp} attempts to learn a mapping from a linear elasticity simulation to its nonlinear counterpart. Common to these methods is that they usually precompute a reduced space or mapping for a {\it specific} rest shape but are able to perform simulations for a wide range of Neumann and Dirichlet boundary constraints. In our case, however, we are facing a significantly different scenario. First, we need to predict and optimize the behavior of a whole range of rest shapes, which are defined by manufacturing feasibility criteria (in our case, close to, but not necessarily perfect, rectangular flat panels). Second, our boundary conditions are fully specified by a low-dimensional boundary curve that corresponds to the attachment frame of the glass panel. Instead, we propose directly infer the deformation and maximal stress from the boundary curve.

Recently, data-driven methods have shown great potential for interactive design space exploration and optimization, for example, for garment design~\cite{wang2019}, or for optimized tactile rendering based on a data-driven skin mechanics model~\cite{verschoor2020tactilerendering}. An overview of graphics-related applications of deep learning can be found in Mitra et al.~\shortcite{Mitra:2019:CDL}. In the context of computational fabrication, data-driven approaches were used, for example, for interactively interpolating the shape and performance of parameterized CAD models~\cite{Schulz2017} or learning the flow for interactive aerodynamic design~\cite{umetani2018learning}. Although these methods are based on an explicit interpolation scheme of close neighbors in the database (\cite{Schulz2017}) or Gaussian processes regression (\cite{umetani2018learning}), in our work, we demonstrate and evaluate the potential of predicting the behavior and solving the inverse problem of designing a cold bent glass façade using neural networks. This entails the additional challenge of dealing with multistable equilibrium configurations that, to the best of our knowledge, has not been addressed before in a data-driven computational design problem.

\section{Overview}

\begin{figure}[t]
    \centering
    \includegraphics{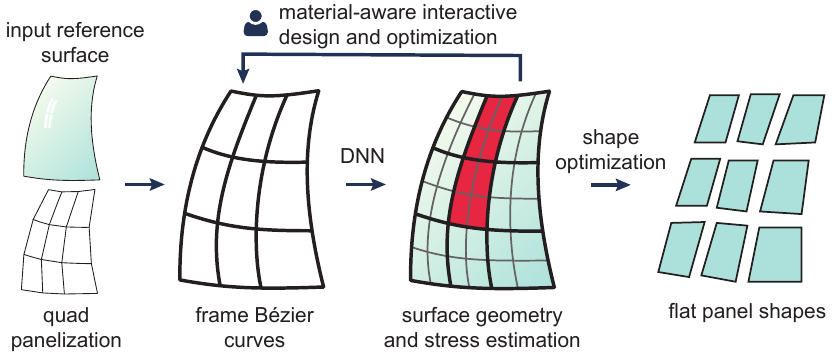}
    \caption{Overview of our design tool workflow. The user makes edits on a quadrilateral base mesh and gets immediate feedback on the deformed shape and maximal stress of the glass panels. When needed, an optimization procedure interactively refines the surface to minimize safety and fairness criteria. If desired, any target reference surface may be used to initialize the process.}
    \label{fig:overview}
\end{figure}

We propose a method for the interactive design of freeform surfaces composed of cold bent glass panels that can be seamlessly integrated in a typical architectural design pipeline. Figure~\ref{fig:overview} shows an overview of the design process. The user makes edits on a base quad mesh that is automatically completed by our system to a mesh with curved B\'ezier boundaries. Our data-driven model then interactively provides the deformed shape of the cold bent glass panels in the form of B\'ezier patches conforming to the patch boundaries and the resulting maximal stress. This form-finding process helps the designer make the necessary decisions to avoid panel failure. At any point during the design session, the user can choose to run our simulation-based optimization method to automatically compute a suitable panelization while retaining some desirable features such as surface smoothness and closeness to the reference design.

In Section~\ref{sec:geometry}, we show how the base mesh controlled by the user is extended through special cubic B\'ezier curves to the set of patch boundaries. Each patch is delimited by planar boundary curves of minimum strain energy. These special B\'ezier patch boundaries are convenient for modeling glass panels because they facilitate the construction of supporting frames while providing a smooth approximation to the desired design.

B\'ezier boundaries do not convey any information on the deformed or undeformed configuration of the panel. Our method uses simulation to compute both configurations of the panel such that certain conditions are met, which are derived from manufacturing constraints. First, current panel assembly does not guarantee $C^1$ continuity at the boundary between neighbor panels because it is very hard to enforce normals along the frame in practice. Second, glass panels have a low tensile strength and are prone to breaking during the installation process in the presence of large tangential forces. Following these criteria, we let the panel be defined by the boundary curve of the frame and compute both the deformed and undeformed shapes of the panel such that the resulting total strain energy is minimal. In this way, we ensure our panelization has at least $C^0$ continuity and that the assembly of the panels requires minimal work, thus reducing the chances of breakage. In Section~\ref{sec:simulation}, we describe in detail the physical model and the computation of minimal energy panels.

Panel shape optimization provides us with a mapping between our design space of B\'ezier boundary curves and theoretically realizable cold bent panels, in both undeformed and deformed configurations. Our material model also accurately estimates the maximum stress endured by the glass. The user is free to interactively edit the base mesh while receiving immediate feedback on the maximum stress, but this neither ensures the panels will not break, nor does it foster the approximation of a target reference surface. To achieve this goal, we solve a design optimization problem: B\'ezier boundary curves are iteratively changed to minimize closeness to an input target surface (and other surface quality criteria) while keeping the maximum stress of each panel within a non-breaking range. In Section~\ref{sec:design}, we describe in detail our formulation of the design optimization.

However, accurately computing the minimal energy panels is computationally very challenging, which makes physical simulation infeasible for being directly used within the design optimization loop. Furthermore, the mechanical behavior of glass panels under compression often leads to multiple stable minimal energy configurations depending on the initial solution. This complicates the optimization even more: not only does the problem turn into a combinatorial one, but there is no algorithmic procedure that can efficiently count and generate all existing static equilibria given some boundary curve. We address this challenge by building a data-driven model of the physical simulation. First, we densely sample the space of the B\'ezier planar boundary curves and compute the corresponding minimal energy glass panels together with an estimation of the maximum stress. Then, we train a Mixture Density Network (MDN) to predict the resulting deformed shape and maximum stress given the boundary of the panel. The MDN explicitly models multistability and also allows us to discover alternative stable equilibria that can be used to enrich the training set. In Section~\ref{sec:learning}, we elaborate on the characteristics of our regression network and our sampling and training method. The trained regression network can finally be used to solve the inverse design optimization problem. Once the user is satisfied with the design, our shape optimization procedure generates the rest planar panels, which are ready to be cut and assembled into a beautiful glass façade.

\section{Geometry representation}
\label{sec:geometry}

A panelization of an architectural surface is built upon a quadrangular base mesh $\mathcal{M} = (V, E, F)$, where the vertices $V$ determine the panel corner points and each quad face in $F$ is filled by one curved panel. In practice, the user interacts with the design tool by making edits to $\mathcal{M}$ through any parametric mesh design method, in our case a Catmull-Clark subdivision from a coarser mesh (see Figure~\ref{fig:teaser} and the supplementary video). This helps achieve fair base meshes and gives a reasonable control for edits. However, any other mesh design scheme could be potentially used.

Each edge in $E$ is then automatically replaced by a planar cubic B\'ezier curve defining the boundaries of the panel, and the inner control points are predicted using our regression model. In this section, we describe the details for getting from $\mathcal{M}$ to the union of curved panels. Moreover, we show how to express the panels with a minimal number of parameters, which are later used for the data-driven model. 

\subsection{Panel parameterization} 
 
We model each glass panel as a bicubic B\'ezier patch $\mathbf{S}:[0, 1]^{2} \rightarrow \RR^{3}$, which is defined by 16 control points $\mathbf{c}_{ij}$, where $i$,$j\in\{0,1,2,3\}$. The corner points $\mathbf{c}_{00}$, $\mathbf{c}_{03}$, $\mathbf{c}_{33}$, $\mathbf{c}_{30}$ are vertices in $\mathcal{M}$. 

\subsubsection{Panel boundary}

\begin{figure}[t]
    \centering
    \begin{overpic}[width=\linewidth]{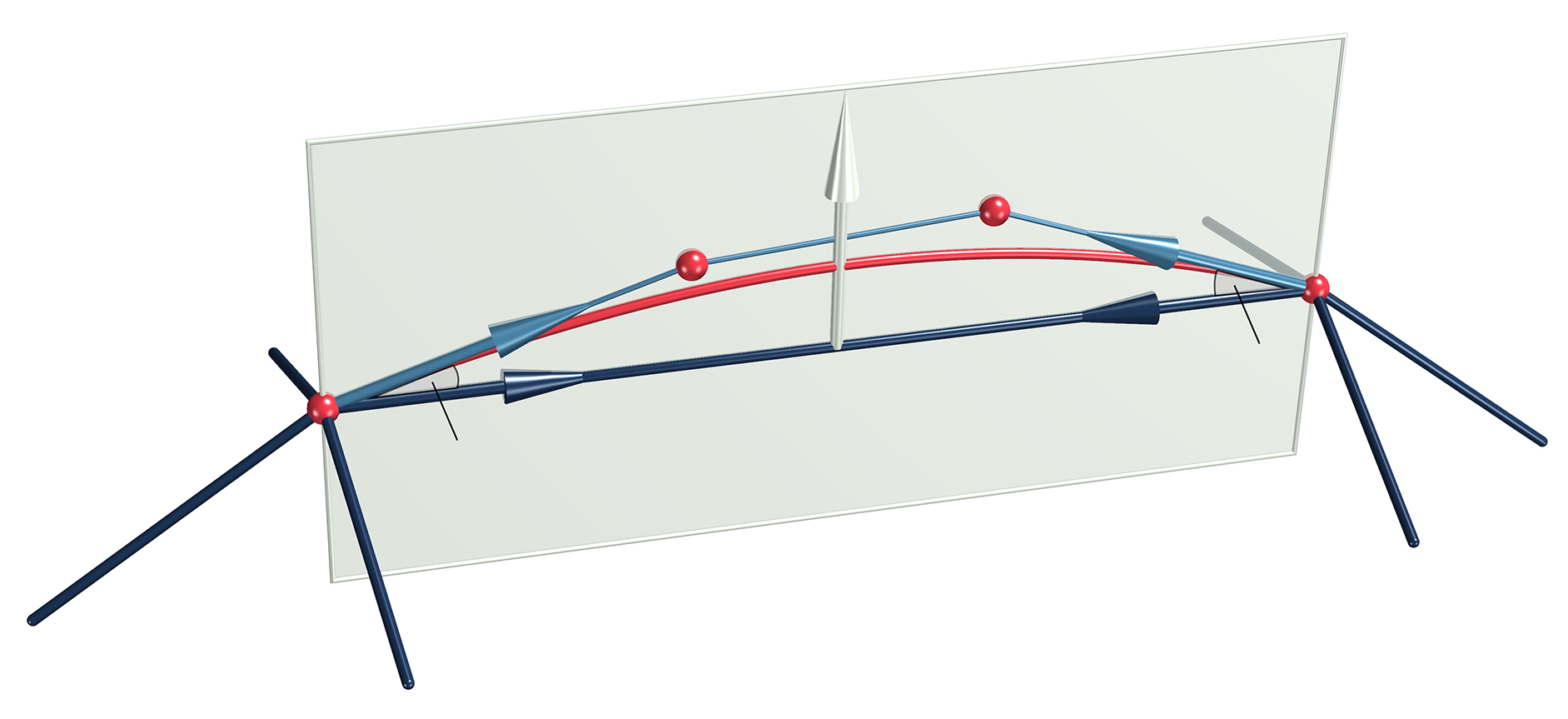}
    \put(15,20){\contour{white}{$\mathbf v_1$}}
    \put(85,28){\contour{white}{$\mathbf v_2$}}
    \put(36.8,31.4){\contour{white}{$\mathbf v_1 \!+\! m_1\mathbf t_1$}}
    \put(56.5,34.6){\contour{white}{$\mathbf v_2  \!+\! m_2\mathbf t_2$}}
    \put(49.5,37){\contour{white}{$\mathbf s_e$}}
    \put(23,33){\contour{white}{$\Pi_e$}}
    \put(31,27){\contour{white}{$\mathbf t_1$}}
    \put(73,32){\contour{white}{$\mathbf t_2$}}
    \put(33.7,17.3){\contour{white}{$\mathbf e_1$}}
    \put(71,22.5){\contour{white}{$\mathbf e_2$}}
    \put(28.7,14.3){\contour{white}{$\theta_1$}}
    \put(80,21){\contour{white}{$\theta_2$}}
    \put(57,26){\contour{white}{$C_e$}}
    \put(53,20){\contour{white}{$e$}}
    \end{overpic}
    \caption{Parameterization of a panel boundary curve from a pair of tangent directions $\mathbf{t}_1$, $\mathbf{t}_2$ corresponding to dual halfedges. The final boundary curve (red) is computed by minimizing a linearized bending energy.}
    \label{fig:panel_boundary}
\end{figure}
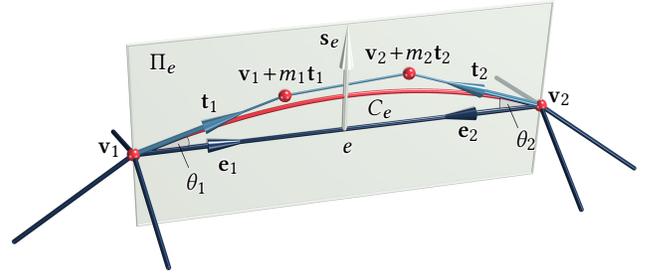

Each edge $e$ of $\mathcal{M}$ is associated with a patch boundary curve $\mathbf{C}_e$. To describe its construction, we focus on a single edge $e$ with vertices $\mathbf{v}_1, \mathbf{v}_2$, and we denote the unit vectors of the half-edges originating at $\mathbf{v}_i$ by $\mathbf{e}_i$ (see Figure~\ref{fig:panel_boundary}). We opted for planar boundary curves representing panel's edges; thus, we first define the plane $\Pi_e$ that contains $\mathbf{C}_e$. We do this by prescribing a unit vector $\mathbf{s}_e\in\RR^3$ that lies in $\Pi_e$ and is orthogonal to $e$. Note that this parameterization is non-injective (vector $-\mathbf{s}_e$ represents the same plane $\Pi_e$), but its ambiguity can be resolved using the compact representation in Section~\ref{subsec:compact-rep}. The two inner control points of the cubic curve $\mathbf{C}_e$ lie on the tangents at its end points. Tangents are defined via the angles $\theta_i$ they form with the edge. Hence, the unit tangent vectors are
%%%%%%%%%%%%%%%
\begin{equation*}
    \mathbf{t}_i = \mathbf{e}_i \cos \theta_i + \mathbf{s}_{e} \sin \theta_i,
\end{equation*}
%%%%%%%%%%%%%%%
In view of our aim to get panels that arise from the flat ones through bending, we further limit the cubic boundary curves to those with a minimal (linearized) bending energy, as described in \cite{YONG2004}. For them, the two inner control points are given by $\mathbf{v}_i + m_i \mathbf{t}_i$, $i=1,2$, with
%%%%%%%%%%%%%%%
\begin{equation*}
    m_1 = \frac
    {(\mathbf{v}_2 - \mathbf{v}_1 ) \cdot [2 \mathbf{t}_1 - ( \mathbf{t}_1 \cdot \mathbf{t}_2) \mathbf{t}_2]
        }
    {4
    -\big( \mathbf{t}_1 \cdot \mathbf{t}_2 \big)^2
    },
\end{equation*}
%%%%%%%%%%%%%%%
and $m_2$ is obtained analogously by switching indices 1 and 2. 

The boundary of $\mathbf{S}$ is thus fully parameterized by the 4 corner vertices $\mathbf{c}_{ij}$, $i,j\in\{0,3\}$, the 4 edge vectors $\mathbf{s}_e$, and the 8 tangent angles $\theta$ (2 per edge). This parameterization of the panels is used in the regression model and the design tool implementation described in Sections~\ref{sec:learning} and \ref{sec:design}, respectively.

\subsubsection{Panel interior}
\label{subsec:panel-param-interior}

The interior control points $\mathbf{c}_{ij}$, $i,j\in\{1,2\}$ express the shape of a panel enclosed by a given boundary. We found that within the admissible ranges of the boundary parameters, any optimal glass panel (see a detailed description in Section~\ref{sec:simulation}) can be very closely approximated by fitting the internal nodes of the B\'ezier patch. Moreover, we need to regularize the fitting, because for a given B\'ezier patch, it is possible to slide the inner control points along its surface while the resulting geometry stays nearly unchanged.

We denote the vertices of the target panel shape $\mathbf{x}_i$ and the corresponding vertex normals $\mathbf{n}_i$. For every $\mathbf{x}_i$, we find the closest points $\mathbf{y}_i$ on the B\'ezier surface and fix their coordinates in the parameter domain. The fitting is then formulated as follows:
%%%%%%%%%%%%%%%
\begin{equation*}
\begin{aligned}
    \min_{\mathbf{c}_{ij}} \quad &
    \big \|
    \big( \mathbf{y}_i(\mathbf{c}_{ij}) - \mathbf{x}_i \big) \cdot \mathbf{n}_i
    \big \|^2 A_i
        + w_\mathrm{B} \sum_k E_k^2(\mathbf{c}_{ij}), \quad
        i,j\in\{1,2\},
\end{aligned}
\end{equation*}
%%%%%%%%%%%%%%%
where $A_i$ are Voronoi cell areas per panel vertex, $E_k$ are the lengths of all control mesh edges incident to the internal nodes, and $w_\mathrm{B}$ is the regularizer weight that we set to $10^{-5}$. To achieve independence of rigid transformations, we express the inner control points in an orthonormal coordinate frame adapted to the boundary. The frame has its origin at the barycenter of the four corner points. Using the two unit diagonal vectors
%%%%%%%%%%%%%%%
\begin{equation*}
    \mathbf{g}_0=\frac
        {\mathbf{c}_{33} - \mathbf{c}_{00}}
        {\| \mathbf{c}_{33} - \mathbf{c}_{00} \|},
    \quad
    \mathbf{g}_1=\frac
        {\mathbf{c}_{30} - \mathbf{c}_{03}}
        {\| \mathbf{c}_{30} - \mathbf{c}_{03} \|},
\end{equation*}
%%%%%%%%%%%%%%%
the $x$-axis and $y$-axis are parallel to the diagonal bisectors, $\mathbf{g}_1 \pm \mathbf{g}_0$, and the $z$-axis is parallel to $\mathbf{b}=\mathbf{g}_0 \times \mathbf{g}_1$, which we call the face normal.

\subsection{Compact representation}
\label{subsec:compact-rep}

The panel boundary is used as an input to a neural network to predict the shape and stress of the minimal energy glass panel(s) conformal to that boundary. Thus, it is beneficial to reduce the input to the essential parameters, eliminating rigid transformations of the boundary geometry. 

We consider $\mathbf{d}\in \RR^6$ to be the vector of the six pairwise squared distances of vertices $\mathbf{c}_{ij}$, $i,j\in\{0,3\}$. Given $\mathbf{d}$, we can recover two valid mirror-symmetric embeddings of the 4 corner points. Assuming that the order of the vertices is always such that 
%%%%%%%%%%%%%%%
\begin{equation*}
    \det( \mathbf{c}_{03} - \mathbf{c}_{00},
        \mathbf{c}_{30} - \mathbf{c}_{00},
        \mathbf{c}_{33} - \mathbf{c}_{00})
    \ge 0
\end{equation*}
%%%%%%%%%%%%%%%
holds, the embedding is unique up to rigid transformations. We assume such a vertex ordering from now on. The plane $\Pi_e$ for each edge is then characterized by its oriented angle $\gamma_e$ with the face normal $\mathbf{b}$. Finally, we define $\mathbf{p}\in \RR^{18}$ as the concatenation of the distance vector $\mathbf{d}$, the 4 edge plane inclinations $\gamma_e$, and the 8 tangent angles $\theta$ (2 per edge). The vector $\mathbf{p}$ is used as an input to the neural network defined in Section~\ref{sec:learning}.

\section{Panel shape optimization}
\label{sec:simulation}

Our method leverages mechanical simulation to create a large dataset of minimal energy panels that conform to cubic B\'ezier boundaries. Given some boundary curves, we are interested in finding deformed glass configurations that are as developable as possible. Non-developable panels result in high tangential forces that complicate the installation of the panel and increase the chances of breakage. By finding the pair of deformed and undeformed shapes of the panel that minimize the strain energy subject to a fixed frame, we ensure the work required for its installation is minimal, helping to reduce the tangential force exerted at the boundary. This dataset is used to train and test a model that predicts the deformed state and maximum stress of such panels, which is suitable for rapid failure detection and inverse design. In this section, we describe the simulation method used for the computation of the deformed and undeformed states of a minimal energy glass panel. 

\subsection{Continuous formulation}

We aim to define a mechanical model that is sufficiently precise to accurately predict glass stresses under small strains, but still suitable for the fast simulation of a very large number of deformation samples. Consequently, we make some reasonable simplifying assumptions in a similar way to Gingold et al.~\shortcite{Gingold2004}. We geometrically represent a glass panel as a planar mid-surface extruded in two opposite normal directions by a magnitude $h/2$, where the total thickness $h$ is much smaller than the minimal radius of curvature of the reference boundary frame. We assume the lines normal to the mid-surface always remain straight and do not undergo any stretching or compression. Under a linearity assumption, the following expression for the volumetrically defined Green's strain tensor $\mathbf{E}$ with offset $z$ in the normal direction can be derived:
%%%%%%%%%%%%%%%
\begin{equation}
    \label{eq:ct_strain}
    \mathbf{E}(\mathbf{x}, \mathbf{\bar x},z) = \mathbf{\bar E}(\mathbf{x},\mathbf{\bar x}) + z \mathbf{\hat E}(\mathbf{x},\mathbf{\bar x}).
\end{equation}
%%%%%%%%%%%%%%%
Here, $\mathbf{x}$ and $\mathbf{\bar x}$ are, respectively, the deformed and undeformed configurations of the mid-surface, and $\mathbf{\hat E}$ is the quadratic bending strain, equivalent to the shape operator of the deformed mid-surface. The membrane strain $\mathbf{\bar E} = 0.5(\mathbf{F}^T\mathbf{F} - \mathbf{I})$ is the in-plane Green's strain tensor defined in terms of the deformation gradient $\mathbf{F}$. We refer to Gingold et al.~\shortcite{Gingold2004} for a detailed explanation of the continuous formulation. We will focus on our discrete formulation, which has been previously considered by Weischedel~\shortcite{Weischedel2012ADG}.

\begin{figure}[t]
    \centering
    \begin{overpic}{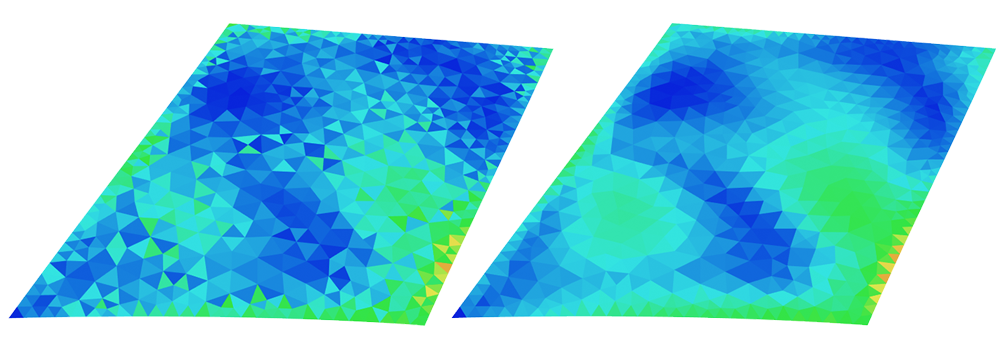}
    \put(0,28){\stressbar}
    \end{overpic}
    \caption{A comparison between the stress distribution produced with the typical shape operator used in, e.g., \cite{pfaff2014cracking} (left), and ours, as suggested in \cite{Grinspun2006} (right). The latter is much smoother and results in a more reliable estimation of the maximal stress.}
    \label{fig:smooth_stress}
\end{figure}

\subsection{Discrete formulation}
We discretize glass panels using a triangulated surface mesh with $N$ nodes and $M$ edges. We separately consider the membrane and bending strains from Equation~(\ref{eq:ct_strain}) and define two corresponding mid-surface energy densities integrated over the panel thickness.

\subsubsection{Membrane energy density}
To discretize membrane strain, we assume piecewise constant strains over FEM elements. In this context, in-plane Green's strain is computed as follows:
%%%%%%%%%%%%%%%
\begin{equation}
    \mathbf{\bar E} = \frac{1}{16 A^2}
    \sum_{i}^3 s_i (\mathbf{t}_j\otimes\mathbf{t}_k + \mathbf{t}_k\otimes\mathbf{t}_j),
    \label{eq:membraneStrain}
\end{equation}
%%%%%%%%%%%%%%%
where $A$ is the triangle area, $s_i = \bar l_i^2 - l_i^2$ ($i$'th edge strain), $\mathbf{t}_j$ and $\mathbf{t}_k$ are the two other edge vectors rotated by $-\pi/2$. For computing the corresponding membrane energy density integrated over the panel thickness, we adopt the Saint Venant-Kirchhoff model:
%%%%%%%%%%%%%%%
\begin{equation}
    \bar W = h \Big(
        \frac{\lambda}{2} \big(\Tr\mathbf{\bar E}\big)^2
        + \mu\Tr\big(\mathbf{\bar E}^2\big)
    \Big),
    \label{eq:membraneEnergy}
\end{equation}
%%%%%%%%%%%%%%%
where $\lambda$ and $\mu$ are, respectively, first and second Lam\'e parameters.

\subsubsection{Bending energy density}
The bending strain is directly defined as the geometric shape operator of the continuous surface. We compute a discrete approximation of the shape operator using the triangle-based discretization suggested in Grinspun et al.~\shortcite{Grinspun2006}, which faithfully estimates bending strain regardless of the irregularity of the underlying triangle mesh. In addition to mesh nodes, this metric considers additional DoFs per edge by defining the deviation of the mid-edge normals from the adjacent triangle-averaged direction:
%%%%%%%%%%%%%%%
\begin{equation}
    \mathbf{\hat E} =
    \sum_{i}^3 \frac{\theta_i / 2 + \phi_i}{A l_i} (\mathbf{t}_i \otimes \mathbf{t}_i).
    \label{eq:bendingStrain}
\end{equation}
%%%%%%%%%%%%%%%
Here, $\theta_i$ is a dihedral angle associated with the edge $i$ and $\phi_i$ is the deviation of the mid-edge normal toward the neighbor triangles normals. Overall, the discrete deformed state of the glass panel is defined with a vector $\mathbf{x} \in \mathcal{R}^{3N+M}$. We denote the corresponding undeformed configuration $\mathbf{\bar x}$. The bending energy density integrated over the panel thickness is then defined by the Koiter's shell model~\shortcite{Koiter1966OnTN}:
%%%%%%%%%%%%%%%
\begin{equation}
    \hat W =
    \frac{\mu h^3}{12}
    \Big(
        \frac{\lambda}{\lambda + 2\mu} \big( \Tr\mathbf{\hat E} \big)^2 + 
        \Tr\big(\mathbf{\hat E}^2\big)
    \Big).
    \label{eq:bendingEnergy}
\end{equation}
%%%%%%%%%%%%%%%
Contrary to the simpler thin shell bending models commonly used in computer graphics \cite{pfaff2014cracking}, the discrete shape operator suggested in ~\cite{Grinspun2006} more faithfully captures principal strain curves and outputs smoother stress distributions (Figure~\ref{fig:smooth_stress}). In the next section, we will describe how we find the minimal energy configuration corresponding to some given B\'ezier boundaries.

\begin{figure}[b]
    \centering
    \includegraphics[width=1\linewidth]{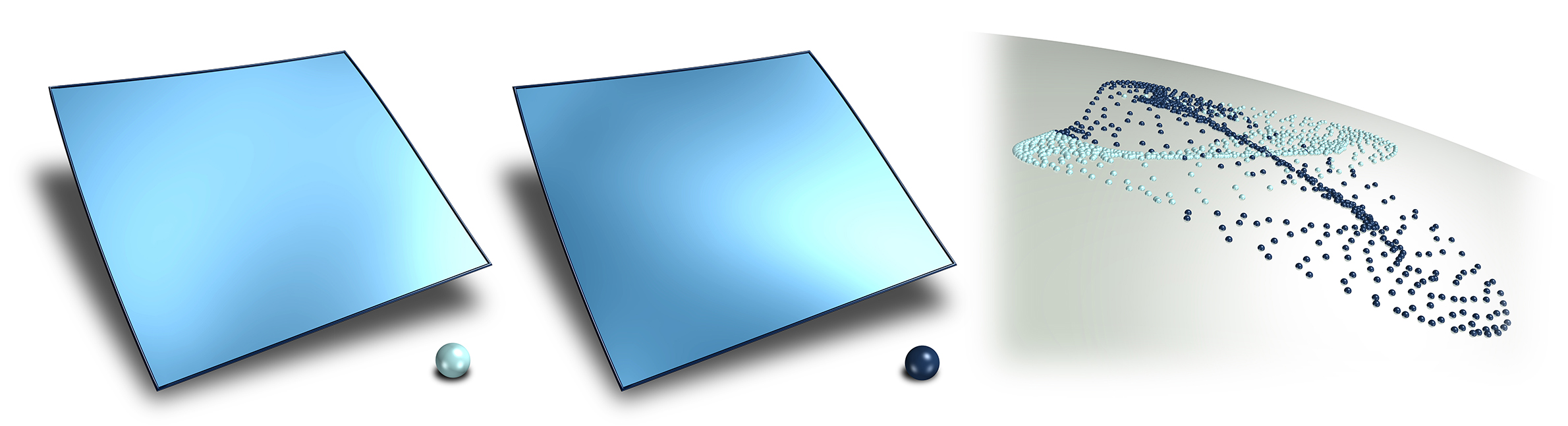}
    \caption{Comparison between two alternative stable equilibria for a given B\'ezier boundary. The two resulting panels produce radically different Gauss maps (right), leading to very distinguishable reflection effects.}
    \label{fig:multistability}
\end{figure}

\subsection{Minimal energy panels}
\label{subsec:minimal_energy_panels}

Given a parametric design of a façade composed of a quadrangular mesh with B\'ezier curves at the edges, we aim to find a suitable panelization using cold bent glass. Although deforming a glass panel to conform to cubic boundaries is feasible, the fragility of this material imposes non-trivial constraints on the maximum amount of stress tolerated by the panels. Thus, we designed a method to compute the glass panel design with the lowest possible strain energy that still will fit our installation constraints. Note that in practice, the existing assembly methods do not preserve normals across neighboring panels; thus, we restrict our problem to guarantee only $C^0$ continuity. By computing the fabricable panel with the lowest possible total strain energy, we minimize the net work required to install the panel and notably reduce the local tangential stress suffered by the glass.

Overall, our pipeline takes as an input the 16 control points of the  B\'ezier boundaries and automatically computes both the deformed $\mathbf{x}$ and undeformed $\mathbf{\bar x}$ configurations of a planar glass panel that conforms to the boundary and has minimal energy. This is done in two steps.

\subsubsection{Initialization} 
At first, we generate a regular mesh that uniformly discretizes the parameter domain of the surface (a unit square) and lift the vertices to an initial B\'ezier patch defined by the boundaries. In our pipeline, such a patch can be obtained in two ways:
\begin{itemize}
    \item Generated by our prediction model, when shape optimization is used to enrich the database or to compute the undeformed shape of the final design panels.
    \item Initialized as a surface patch with zero twist vectors at the corners (a quad control mesh has parallelograms as corner faces) when shape optimization is used to build the initial database.
\end{itemize}
The lifted mesh is conformally flattened with minimal distortion. We uniformly resample the boundary of this mesh targeting a total number of edges $M_{\rm{b}}$ and triangulate the interior using Delaunay triangulation with the bounded maximal triangle area. Finally, the vertices of this mesh are mapped back to the parameter domain and lifted to the initial B\'ezier patch. As a result, we obtain an initial configuration for a deformed glass panel conforming to B\'ezier boundaries and its corresponding undeformed configuration.

\subsubsection{Minimization}

The initial solution is not in static equilibrium and has arbitrarily high stresses. We compute the minimal energy configuration by minimizing the discrete strain energies defined in Equations~\ref{eq:membraneEnergy} and \ref{eq:bendingEnergy} over deformed $\mathbf{x}$ and undeformed $\mathbf{\bar x}$ configurations. We refer to the vector of all the deformed nodes at the boundary and the internal nodes as $\mathbf{b}$ and $\mathbf{i}$, respectively. To reduce the complexity of the problem and keep a high-quality triangulation of the undeformed configuration, we assume internal nodes at the rest configuration $\mathbf{\bar i}$ are computed through Laplacian smoothing of the boundary vertices $\mathbf{\bar i} = \mathbf{L}\mathbf{\bar b}$. Then, the aforementioned minimization problem results in the following: 
%%%%%%%%%%%%%%%
\begin{equation}
    \min_{\mathbf{i},\mathbf{\phi}, \mathbf{\bar b}} W(\mathbf{x}, \mathbf{\phi}, \mathbf{\bar b}) + R(\mathbf{\bar b}),
    \label{eq:simulation}
\end{equation}
%%%%%%%%%%%%%%%
where $W$ is the sum of all strain energy terms, $\phi$ are the mid-edge normal deviations, and $R$ is a regularization term removing the null space due to the translation and rotation of the undeformed configuration. In particular, it is formulated as a soft constraint: the centroid of the boundary nodes is fixed to the origin, and one of the nodes has a fixed angle with the x-axis. Note that we only consider undeformed boundary nodes $\mathbf{\bar b}$ as DoFs of the optimization; after each solver iteration, we project the internal nodes' coordinates $\mathbf{\bar i}$ through Laplacian smoothing. In addition, the boundary nodes of the deformed configuration remain fixed and conforming to B\'ezier boundaries.

As can be seen in Figure~\ref{fig:multistability}, minimizing Equation~\ref{eq:simulation} does not always produce a unique solution. For a given boundary, glass panels can potentially adopt multiple stable equilibria corresponding to locally optimal shapes that depend on the initialization of the problem. Although for some boundary curves there is a clearly preferred shape that is more energetically stable than the rest, in other cases, several stable equilibria are valid solutions that might be practically used in a feasible panelization. Furthermore, the maximum stress levels differ a lot between stable configurations. Multistability imposes two challenges for building a data-driven model of glass panel mechanics. First, we do not know in advance the number of local minima that exist for a given boundary nor how energetically stable these configurations are in practice; second, we do not know how to initialize the minimization problem to obtain such solutions. Both challenges motivated the use of a MDN as a regressor for the shape and corresponding stress of the glass panels. In Section~\ref{sec:learning}, we describe our regression model and the methodology we followed to enrich the database by discovering new stable equilibria through an iterative process. 

\subsection{Failure criterion}

To estimate whether the panel is going to break, we compute the maximal engineering stress across all the elements of the discretization. We estimate the stress of an element by computing the first Piola-Kirchhoff stress tensor $\mathbf{P} = \mathbf{F}\mathbf{S}$. Here, $\mathbf{F}$ is the deformation gradient of the element, and $\mathbf{S}$ is the corresponding second Piola-Kirchhoff stress tensor. In a similar fashion to Pfaff et al.~\shortcite{pfaff2014cracking}, we compute the total stress of a panel using our estimation of the combined bending and membrane strain introduced in Equation~\ref{eq:ct_strain}:
%%%%%%%%%%%%%%%
\begin{equation}
    \mathbf{S}\big(\mathbf{E}(\mathbf{\bar E}, \mathbf{\hat E}, z)\big) =
    \lambda \Tr(\mathbf{E}) \mathbf{I} + 2 \mu \mathbf{E}.
    \label{eq:second_piola}
\end{equation}
%%%%%%%%%%%%%%%

The maximal engineering stress is then evaluated as the maximum absolute singular value of $\mathbf{P}$ across all elements. That is, for each element, we compute $\mathbf{S}\big(\mathbf{E}(\mathbf{\bar E}, \mathbf{\hat E}, \pm h/2)\big)$, where the bending contribution to the stress is at its maximum, and pick the highest absolute singular value. The global maximal stress value is generally at most $C^0$-continuous with respect to the panel boundary curves, which makes its direct usage in a continuous optimization undesired. Instead, we compute an $L_p$-norm of maximal principal stress per element. In practice, we found that $p = 12$ suffices. We denote the resulting value $\sigma$ and refer to it as the ``maximal stress'' for brevity.

Taking our assumptions, it is important to note that neither the overall shape nor the maximal stress value changes for a given panel under uniform scaling. This implies that only the ratio of the thin shell dimensions and the panel thickness matters. For simplicity, we choose 1~mm as our canonical thickness for the simulations and scale the obtained results accordingly for every other target thickness.

\section{Data-driven model}
\label{sec:learning}

We require a model that can efficiently predict the shape and stress of the minimum-energy panels for a given boundary. The simulation described in Section~\ref{sec:simulation} calculates these quantities, but is too slow to incorporate in an interactive design tool. Our data-driven model aims to predict the deformed shape and corresponding maximum stress of the panels more efficiently. Moreover, we use it to calculate the derivatives with respect to the input boundary, which is required for gradient-based design optimization.

Therefore, we use a statistical model that maps panel boundaries to the shapes and stresses of minimal-energy conforming cold bent glass surfaces. Section~\ref{sec:ddm-pred-model} describes the model and training process. The training requires a large dataset of boundaries and the resulting panel shapes and stresses; in Section~\ref{sec:ddm-initial-dataset}, we describe the space of boundaries we sample from and how the shape optimization and stress computation of Section~\ref{sec:simulation} is applied to them. To improve the results further, we augment the dataset to better cover the regions of the input space where the predictions do not match the training data because of multistability of the glass panels (Section~\ref{sec:ddm-dataset-enrichment}) and retrain the model on this enriched dataset. We will release our dataset and pretrained model publicly.

\subsection{Multi-modal regression model}
\label{sec:ddm-pred-model}

Our prediction model takes as an input a vector $\mathbf{p} \in \RR^{18}$ representing a panel boundary. As noted in Section~\ref{subsec:minimal_energy_panels}, several different surfaces may conform to a given boundary, corresponding to different local minima of the strain energy. Therefore, predicting a single output yields poor results, typically the average over possible shapes. Instead, we use a \textit{mixture density network} (MDN)---a neural network model with an explicitly multi-modal output distribution~\cite{bishop06book}. For a given boundary, each mode of this distribution should correspond to a different conforming surface.

Whereas training a neural network to minimize the mean squared error is equivalent to maximizing the data likelihood under a Gaussian output distribution, an MDN instead maximizes the likelihood under a Gaussian mixture model (GMM) parameterized by the network. Therefore, it must output the means and variances of a fixed number $K$ of mixture components, as well as a vector $\bm{\hat{\pi}}$ of component probabilities. In the rest of the paper, all variables with hats denote predictions from our data-driven model, as opposed to values from the physical simulation. In our model, each component is a (12 + 1)-dimensional Gaussian with diagonal covariance, corresponding to the four interior control points of the shape, $\mathbf{c}_{ij}\in \RR^3$, $i,j\in\{1,2\}$, and stress, $\sigma$, of one possible conforming surface. We denote the mean of the concatenated shape and stress of the $k$\textsuperscript{th} mixture component by $\bm{\hat{\zeta}}^k$ and the variance by $\bm{\hat{\xi}}^k$; both are output by the neural network and hence depend on the input boundary $\mathbf{p}$ and network weights $\mathbf{w}$.

\begin{figure}[t]
    \centering
    \includegraphics{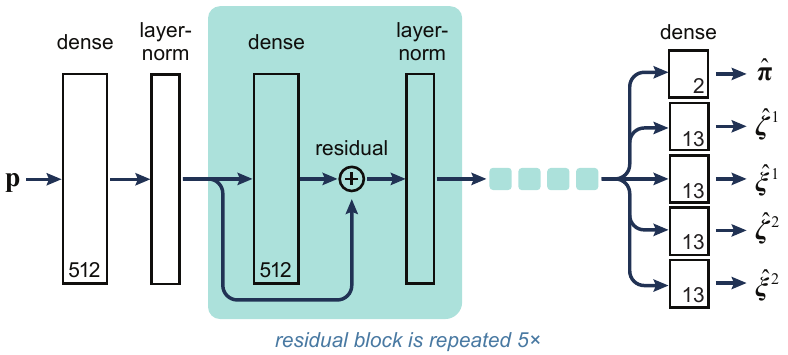}
    \caption{
    Architecture of our data-driven model. The input is a panel boundary $\mathbf{p}$; the model predicts means $\bm{\hat{\zeta}}^k$, variances $\bm{\hat{\xi}}^k$, and component weights $\bm{\hat{\pi}}$ for a two-component Gaussian mixture over the shape and stress of the minimal-energy surface.
    The numbers in dense layers indicate the number of output units.
    }
    \label{fig:dnn}
\end{figure}

\subsubsection{Model architecture}

We use a densely connected model with six layers of 512 exponential linear units (ELU)~\cite{clevert15arxiv}, with residual connections~\cite{he16cvpr} and layer-normalization~\cite{ba16arxiv} at each hidden layer (Figure~\ref{fig:dnn}). We trained tens of models using combinations of these hyperparameters' values and selected the one that performed the best on the held-out validation set. In simulation, we observed that a given boundary could potentially admit more than two stable states. However, these cases were extremely rare; therefore, we set $K = 2$. This suffices for capturing the vast majority of stable states observed in our dataset, resulting in a low validation error. Hence, the output layer has 54 units, with no activation for the means $\bm{\hat{\zeta}}^k$, exponential activation for the variances $\bm{\hat{\xi}}^k$, and a softmax taken over the mixing probabilities $\bm{\hat{\pi}}$.

\subsubsection{Model training}

The model is trained to minimize the negative log-likelihood of a training set $\mathcal{T}$ under the GMM:
\begingroup\makeatletter\def\f@size{8.25}\check@mathfonts
\begin{equation}
    \hspace{-1pt} \mathcal{L} \left( \mathcal{T}; \mathbf{w} \right) = 
    - \hspace{-1pt}
    \sum_{(\mathbf{p}, \bm{\zeta}) \in \mathcal{T}}
    \log \left\{ 
        \sum_{k=1}^K 
        \hat{\pi}^k(\mathbf{p} ; \mathbf{w}) \, 
        \mathcal{N} \hspace{-1.5pt} \left( \bm{\zeta} \, \big| \, 
            \bm{\hat{\zeta}}^k ( \mathbf{p} ; \mathbf{w} ) ,\, 
            \bm{\hat{\xi}}^k ( \mathbf{p} ; \mathbf{w} ) 
        \right)
    \right\}
\end{equation}
\endgroup
where $\bm{\zeta}$ is a true output for panel $\mathbf{p}$, i.e. the concatenation of shape and stress from one simulation run, and $\mathcal{N}$ represents a diagonal Gaussian density. We also add an L2 regularization term with strength $10^{-4}$ on the weights $\mathbf{w}$, to discourage over-fitting.

We use the stochastic gradient method Adam \cite{kingma15iclr} to minimize the above loss function with respect to the network weights $\mathbf{w}$. We use a batch size of 2048, learning rate of $10^{-4}$, and early stopping on a validation set with patience of 400 epochs. We select the best model in terms of the validation loss obtained during the training process. A single epoch takes approximately 30 seconds on a single NVIDIA Titan X graphics card, and in total, training takes around 20 hours.

\subsubsection{Model output}

For brevity, in the remainder of the paper, for a given panel boundary $\mathbf{p}$ and for a possible state $k \in \{1,2\}$, we write:
\begin{itemize}
    \item $\hat{\mathbf{S}}_{\mathbf{p}}^k : \left[ 0, 1 \right]^2 \rightarrow \RR^3$ for the B\'ezier surface patch that is defined by the boundary $\mathbf{p}$ and the predicted interior control nodes from the mean of the $k$\textsuperscript{th} component (i.e., the leading 12 elements of $\bm{\hat{\zeta}}^k$).
    
    \item $\hat{\sigma}_{\mathbf{p}}^k$ for the stress value (i.e., the last element of $\bm{\hat{\zeta}}^k$).
    
    \item $\hat{\pi}_\mathbf{p}^k$ for the $k$\textsuperscript{th} component probability $\hat{\pi}^k(\mathbf{p} ;\, \mathbf{w})$.
\end{itemize}

Furthermore, we write $\hat{\mathbf{S}}_{\mathbf{p}}$ and $\hat{\sigma}_{\mathbf{p}}$ (i.e., without the $k$ superscript) to refer to the \textit{best} prediction for boundary $\mathbf{p}$, which is determined by two factors:
\begin{enumerate}
    \item if any of the component probabilities $\hat{\pi}_\mathbf{p}^k$ is greater than 95\%, we discard the alternative and define the corresponding shape/stress prediction as best, or
    
    \item otherwise, we consider both components as valid, and the best one is determined depending on the application, either as the lower stress, the smoother shape, or the closer shape to a reference surface.
\end{enumerate}
We discard components with $\hat{\pi}_\mathbf{p}^k<0.05$ since the modes with a near-zero probability imply a low level of confidence in the corresponding prediction.

\begin{figure*}[t!]
    \begin{overpic}[width=1\linewidth]{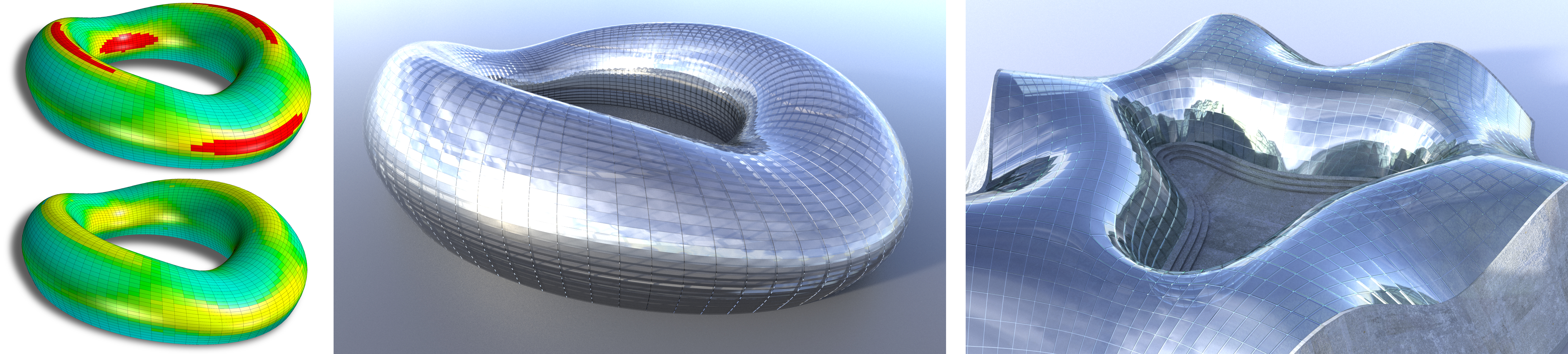}
    \put(0,-1.5){\stressbar}
    \put(0,10){b}
    \put(0,21){a}
    \end{overpic}
    \caption{An initial design includes panels exceeding stress limits (left, a). It is optimized for stress-reduction (left, b) and rendered (center). Right: a different façade designed with our tool.}
    \label{fig:ring}
\end{figure*}

\subsection{Dataset construction}
\label{sec:ddm-initial-dataset}

To train our prediction model, we require a dataset of boundaries that is representative of our target application. These are then paired with the shapes and stresses of the conforming surfaces with minimal energy. Recall from Section~\ref{sec:geometry} that a panel boundary may be parameterized invariantly to rigid transformations by corner pairwise squared-distances $\bm{d}$, edge-plane inclinations $\bm\gamma$, and halfedge tangent directions $\bm\theta$. We generate boundaries by sampling these parameters from the ranges and distributions described in Appendix~\ref{app:boundary-sampling}. Note that the physical model for the deformed shape and stress is invariant under the scaling of all geometric magnitudes; we choose our sampling ranges so that it would be possible to scale the results to panel length-to-thickness ratios commonly used in cold bent glass façades (e.g., 150--600). By applying the shape optimization described in Section~\ref{sec:simulation} to these boundaries, we obtain fine discrete meshes representing the deformed cold bent panels. We obtain a B\'ezier representation of such panels by keeping the sampled boundaries and fitting interior control points to match the simulated surface using the method described in Section~\ref{subsec:panel-param-interior}. Plus, in our representation, any non-flat panel geometry can be equivalently represented in four alternative ways, depending on vertex indexing, and can be mirrored. Therefore, we transform each simulated panel into eight samples by permuting the vertex indexes and adding their mirror-symmetric representations.

In total, we simulated approximately 1.5 million panels, which corresponds to 12 million samples after vertex permutations and adding mirror-symmetric panels.
We reserved 10\% of these samples as a validation set for tuning the optimization hyperparameters and network architecture. To acquire such large amounts of data requiring massive computations, we employed cloud computing.

\subsection{Dataset enrichment}
\label{sec:ddm-dataset-enrichment}

When a given boundary has multiple conforming panels, the physical simulation returns only one of these determined by the twist-free B\'ezier patch initialization. Conversely, our data-driven model always predicts $K = 2$ states, though one may have a very small mixture weight $\hat{\pi}^k$, indicating it is unlikely to be a valid optimal panel. We observed that after training, the model often predicts shapes for boundaries in its training set that differ from those returned by the simulation---however, re-simulating these boundaries with a different initialization recovers a solution close to that predicted by the data-driven model. This observation suggests a method to extend the dataset with new samples to improve prediction error.

Specifically, we use the prediction from the model as an initialization for the simulator, which is then likely to converge to a stable surface that was not reached from the default initialization. The resulting surface can be added to the training set, so after retraining, the model will give an even more accurate prediction in the same region of parameter space. We apply the data-driven model to every panel in the training set, and collect the predicted shapes $\hat{S}^k_{\mathbf{p}}, \, k \in \{1, 2\}$, where $\pi^k > 5\%$. For each of these, we calculate the maximum deviation of the internal control nodes along any dimension, from the true shape in the training set. We then retain the 200k panels ($\sim$15\% of the original training set) for which this deviation is the largest. For each such boundary, we re-run the simulation, using the predicted shape $\hat{S}^k_{\mathbf{p}}$ as the initialization. Finally, we select all panels which have at least 2~mm difference along any dimension of any internal control node compared to the panel obtained originally and add these to the training set. The resulting, enriched training set is used to retrain the model.

\section{Interactive design}
\label{sec:design}

In this section, we show how we arrive at a practical interactive design tool for freeform surface panelization using cold bent glass panels. We aim to produce a tool compatible with the standard design workflow of an architectural designer. At every moment during the editing process, the user gets immediate feedback on the physical properties of the panelization (i.e., shape and stress predictions for the panels). Upon request, an automated process running at interactive rates uses an optimization to ``guide'' the design. Figures~\ref{fig:teaser} and \ref{fig:ring} show two different doubly curved glass surfaces that have been interactively designed from scratch using our tool. Although it is generally desired to create designs free from breaking panels, in a real project, one might like to assume the cost of hot-bending a small proportion of the panels. Therefore, there is a practical trade-off between the smoothness and aesthetics of a design and its manufacturability. We consider this option by explicitly weighting various design criteria in the formulation of our inverse design problem.

\subsection{Optimization formulation}

Depending on the specific application domain, the desired properties might vary. This translates into the minimization of a composite target functional $\mathcal{E}$:
%%%%%%%%%%%%%%%
\begin{equation}\label{eq:func_total}
    \mathcal{E} =
    w_{\sigma} \mathcal{E}_{\sigma} +
     w_{\text{s}} \mathcal{E}_{\text{s}} +
    w_{\text{f}} \mathcal{E}_{\text{f}} +
    w_{\text{p}} \mathcal{E}_{\text{p}} +
     w_{\text{c}} \mathcal{E}_{\text{c}}.
\end{equation}
%%%%%%%%%%%%%%%
Overall, the total energy $\mathcal{E}$ depends on the vertex positions $\mathbf{v}\in V$, the edge plane vectors $\mathbf{s}_e$ defining the plane $\Pi_e$ associated with $e$, the tangent angles $\theta_i$, and some auxiliary variables associated with inequality constraints. Each weighted contribution to $\mathcal{E}$ represents a desired property of the final design, which we discuss in detail in the following sections. 

\subsubsection{Panel stress $\mathcal{E}_{\sigma}$}

The most important property is the manufacturability of the final design. Failure in a specific type of glass is modeled by estimating the maximal stress present in the glass panel and comparing it to the maximum allowed stress value.

The MDN from Section~\ref{sec:learning} acts as a stress estimator. We constrain the predicted stress value $\hat{\sigma}_{\mathbf{p}}$ for a given boundary $\mathbf{p}$ to be less than a stress bound $\sigma_{\max}$. We assign $\sigma_{\max}$ to a value lower than the stress value at which failure occurs, taking into account a safety factor and the estimator error. The inequality constraints $\hat{\sigma}_{\mathbf{p}} \leq \sigma_{\max}$ are converted to equality constraints by introducing an auxiliary variable $u_{\mathbf{p}}\in \RR$ per panel boundary $\mathbf{p}$, and formulating the manufacturability energy as
%%%%%%%%%%%%%%%
\begin{equation}\label{eq:func_stress}
    \mathcal{E}_{\sigma} = \sum_{\mathbf{p}} (\hat{\sigma}_{\mathbf{p}} - \sigma_{\max} + u_{\mathbf{p}}^2)^2.
\end{equation}
%%%%%%%%%%%%%%%
Figure~\ref{fig:lilium_stress45_comp} shows the effect of limiting the maximum stress of the design for a section of the façade of the Lilium Tower.

%-------------------------------------------------------------------
\begin{figure}[t!]
    \centering
    \begin{overpic}[width=1\linewidth]{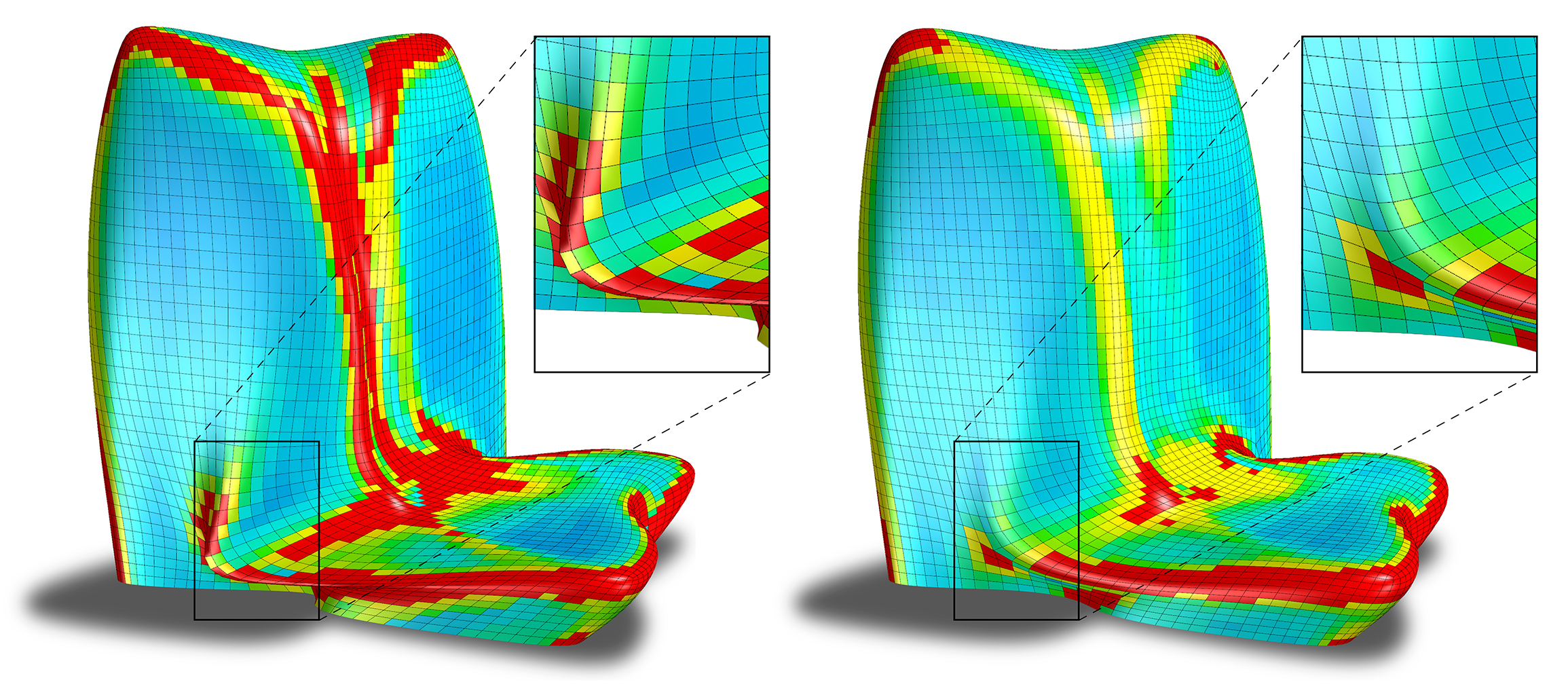}
    \put(0,-2){\stressbar}
    \put(4,42){a}
    \put(53,42){b}
    \end{overpic}
    \caption{Optimization of the NHHQ skyscraper design (Zaha Hadid Architects). We first optimize for smoothness of the overall design, and then optimize selected high stress areas for stress reduction.
    (a) Stress on panels computed on the original shape and panel layout. Red panels exceed the threshold of $65$ MPa. 
    (b) Stress on panels after optimization. The inset shows an area with clearly visible shape change. We decrease the number of panels exceeding $65$ MPa from $1517$ to $874$. }
    \label{fig:nhhq_stress}
\end{figure}
%-----------------------------------------------------------------

%-------------------------------------------------------------------
\begin{figure}[b!]
    \centering
    \includegraphics[width=0.7\linewidth]{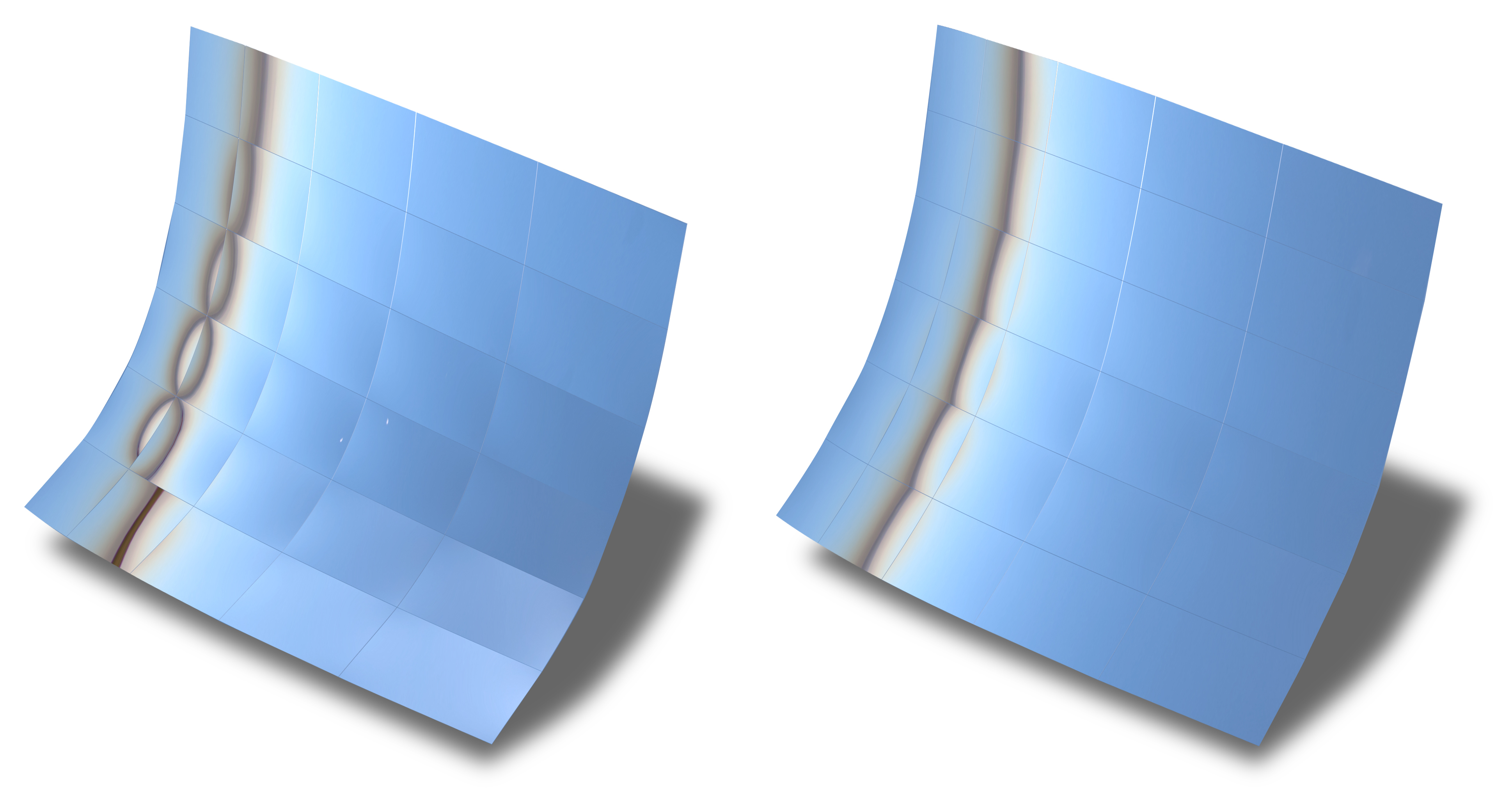}
    \caption{Effect of optimization on visual smoothness. On the left, a selection of cold bent panels computed on a given layout. On the right, the same panels after optimization of the layout for the kink angle and bending stress reduction.}
    \label{fig:square_render_comp}
\end{figure}
%-----------------------------------------------------------------

\subsubsection{Smoothness  $\mathcal{E}_{\text{s}}$.} Here, we collect some terms in the final objective function that aim in various ways to obtain as smooth as possible panelizations. As shown in Figure~\ref{fig:square_render_comp}, this is essential for achieving the stunning look of curved glass façades because it greatly affects the reflection pattern. The smoothness term is the sum of two individual functionals, i.e. $\mathcal{E}_{\text{s}} =  \mathcal{E}_1 + \mathcal{E}_2$.

\subsubsection*{Kink angle smoothing $\mathcal{E}_1$.} 
\begin{figure}[t!]
    \centering
    \begin{overpic}[width=1\linewidth]{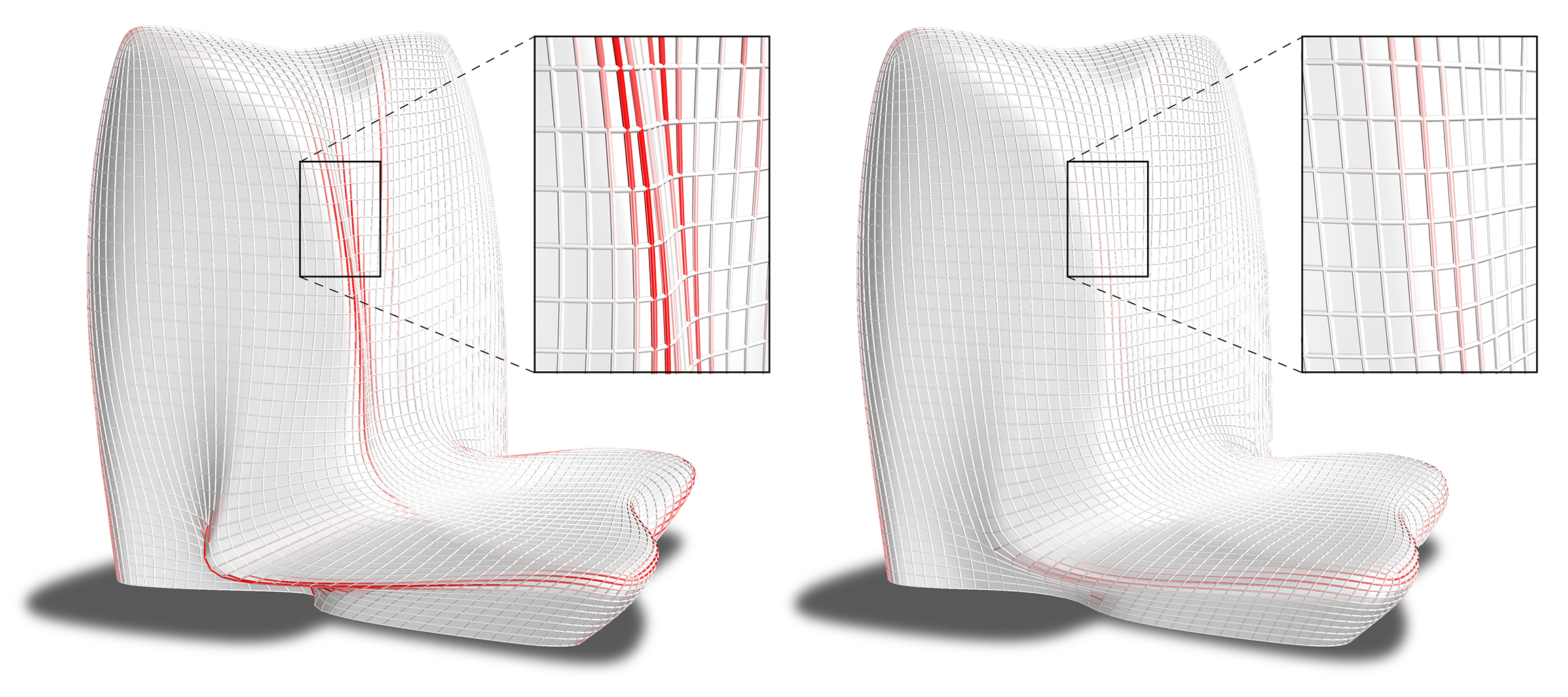}
    \put(0,-2){\kinkbar}
    \put(4,42){a}
    \put(53,42){b}
    \end{overpic}
    \caption{Comparison of the kink angle between panels in the NHHQ model, before (a) and after (b) running our design optimization algorithm.
    As a result, the mean kink angle is lowered from $3.7^\circ$ to $2.7^\circ$, while the maximum is reduced from $60.9^\circ$ to $36.0^\circ$.}
    \label{fig:nhhq_kink}
\end{figure}

It is generally not possible to get smooth connections along the common boundary curves of panels, but we can try to minimize the kink angle. For each pair of faces $f_i$, $f_j$ sharing a common edge $e$, we consider their respective predicted panels $\hat{\mathbf{S}}_i$, $\hat{\mathbf{S}}_j$ and minimize the angle between their surface normals $\mathbf{n}_i$, $\mathbf{n}_j$ evaluated at the parameter $t=0.5$ of the shared curve
%%%%%%%%%%%%%%%
\begin{equation}\label{eq:func_kink_angles}
    \mathcal{E}_{1} = 0.1\sum_{e\in E_{I}} (1 - \mathbf{n}_i \cdot \mathbf{n}_j)^2,
\end{equation}
%%%%%%%%%%%%%%%
where $E_I$ is the set of interior edges of $\mathcal{M}$, and $0.1$ is a suitable importance weight within the smoothness term. Note that $\hat{\mathbf{S}}_i$, $\hat{\mathbf{S}}_j$ are shape predictions for the respective boundary curves of the two faces $f_i$, $f_j$. Thus, optimization involves computing the Jacobian of the MDN output w.r.t. the input boundaries. Figure~\ref{fig:nhhq_kink} shows the effect of including the kink smoothing term in the design of the NHHQ façade.

\subsubsection*{Curve network smoothing $\mathcal{E}_2$.} Each edge in the dominant mesh polylines of $\mathcal{M}$ determines a 
cubic patch boundary curve, and the sequence of these curves should also be as smooth as possible. At each connection of two edges, the corresponding tangents should agree, and thus, the inwards directed unit tangent vectors
satisfy $\mathbf{t}_i=-\mathbf{t}_j$, or equivalently $ \mathbf{t}_{i} \cdot \mathbf{t}_{j} + 1=0$. This
tangent continuity constraint explains the first part in the smoothness term
%
%%%%%%%%%%%%%%%
\begin{equation} \label{eq:smooth}
    \mathcal{E}_{\text{2}} = \sum (\mathbf{t}_{i} \cdot \mathbf{t}_{j} + 1)^2 + 
    \sum \left[ \mathbf{s}_e \cdot ( \mathbf{n}_i + \mathbf{n}_{i+1}) \right]^2.
\end{equation}
%%%%%%%%%%%%%%
The second part concerns the planes $\Pi_e$. We consider an edge $e$ with endpoints $\mathbf{v}_{i}$, $\mathbf{v}_{i+1}$.
The discrete osculating plane at $\mathbf{v}_{i}$ is 
spanned by  $(\mathbf{v}_{i-1}, \mathbf{v}_{i}, \mathbf{v}_{i+1})$ and has a
unit normal $\mathbf{n}_i$. Likewise, $(\mathbf{v}_{i}, \mathbf{v}_{i+1}, \mathbf{v}_{i+2})$ defines a discrete osculating plane
with normal $\mathbf{n}_{i+1}$ at  $\mathbf{v}_{i+1}$.  We want $\Pi_e$ to be the bisecting plane between these two,
i.e. $ \mathbf{s}_e \cdot ( \mathbf{n}_i + \mathbf{n}_{i+1})=0$. Of course, the sums are taken over all occurrences of the described situations. 

Finally, in practice, a few other parts are added to the smoothness term $\mathcal{E}_{\text{s}}$, which concern special cases. At combinatorially singular vertices of $\mathcal{M}$, we constrain the tangent vectors to lie in a tangent plane. Plus, there are various symmetry considerations that are used at the boundary, but those could easily be replaced by other terms with a similar effect. 

%-------------------------------------------------------------------
%-------------------------------------------------------------------

\subsubsection{Mesh fairness $\mathcal{E}_{\text{f}}$}
So far, we have dealt with the smoothness of the panelization to a given mesh $\mathcal{M}$. Because we also allow the mesh $\mathcal{M}$ to change during the design, we need to care about its fairness. This is done in the standard way using second-order differences of consecutive vertices along dominant mesh polylines,
%%%%%%%%%%%%%%%
\begin{equation}\label{eq:func_fair}
    \mathcal{E}_{\text{f}} = \sum (\mathbf{v}_{i-1} - 2\mathbf{v}_{i} + \mathbf{v}_{i+1})^2.
\end{equation}
%%%%%%%%%%%%%%%

\subsubsection{Proximity to reference mesh $\mathcal{E}_{\text{p}}$.}
When designing a panelization for a given reference geometry, it is not sufficient to have the mesh $\mathcal{M}$. One will usually have a finer mesh $\mathcal{M}_{\text{ref}}$ describing the reference geometry (Figure~\ref{fig:overview}). To let $\mathcal{M}$ change but stay close to the reference surface, we need a term that allows for the gliding of $\mathcal{M}$ along $\mathcal{M}_{\text{ref}}$. This is done in a familiar way: to let a vertex  $\mathbf{v_i}$ stay close to $\mathcal{M}_{\text{ref}}$, we consider its closest point $\mathbf{v}_i^*$ on $\mathcal{M}_{\text{ref}}$ and the unit surface normal $\mathbf{n}_i^*$ at $\mathbf{v}_i^*$. In the next iteration, $\mathbf{v_i}$ shall stay close to the tangent plane at $\mathbf{v}_i^*$, which is expressed via
%%%%%%%%%%%%%%%
\begin{equation}\label{eq:func_close}
    \mathcal{E}_{\text{p}} = \sum_{\mathbf{v}_i \in V} \left[ (\mathbf{v_i} - \mathbf{v}_i^*) \cdot \mathbf{n}_i^*\right]^2.
\end{equation}
%%%%%%%%%%%%%%%

\subsubsection{Design space constraints $\mathcal{E}_{\text{c}}$.}
Since we want the neural network to produce reliable estimates, we need to ensure the panel boundary curves remain within the range used for training (Section~\ref{sec:ddm-initial-dataset}). This is achieved as the sum of two constraint functionals $\mathcal{E}_{\text{c}} = \mathcal{E}_3 + \mathcal{E}_4$. First, we constrain the tangent angles to $\abs{\theta_i} = \angle (\mathbf{t}_i, \mathbf{e}_i)\leq 4.9^\circ$ for all angles $\theta_i$ of halfedges $\mathbf{e}_i$ with tangent vectors $\mathbf{t}_i$. We again convert the inequality constraints to equality constraints by introducing auxiliary variables $u_i$,
%%%%%%%%%%%%%%%
\begin{equation}\label{eq:tangent_angle_energy}
    \mathcal{E}_3 =  \sum (\theta_i^2 - (4.9^\circ)^2 + u_i^2)^2.
\end{equation}
%%%%%%%%%%%%%%%
Second, we are working under the assumption that the vectors $\mathbf{s}_e$ are unitary and orthogonal to their respective edges $e$, which results in
%%%%%%%%%%%%%%%
\begin{equation}\label{eq:func_edge_planes_2}
    \mathcal{E}_4 = 
    \sum_e [( \mathbf{s}_e \cdot \mathbf{e})^2 + 
    ( \mathbf{s}_e^2 - 1)^2].
\end{equation}
%%%%%%%%%%%%%%%

\subsection{Optimization solution} \label{ssub:numerical_optimization_algorithm}

The minimization of $\mathcal{E}$ results in a nonlinear least-squares problem that we solve using a standard Gauss-Newton method. The derivatives are computed analytically, and, since each distinct term of $\mathcal{E}$ has local support, the linear system to be solved at each iteration is sparse. We employ Levenberg-Marquardt regularization and sparse Cholesky factorization using the TAUCS library \cite{taucs}.

\begin{figure}[t!]
    \centering
    \begin{overpic}[width=\linewidth]{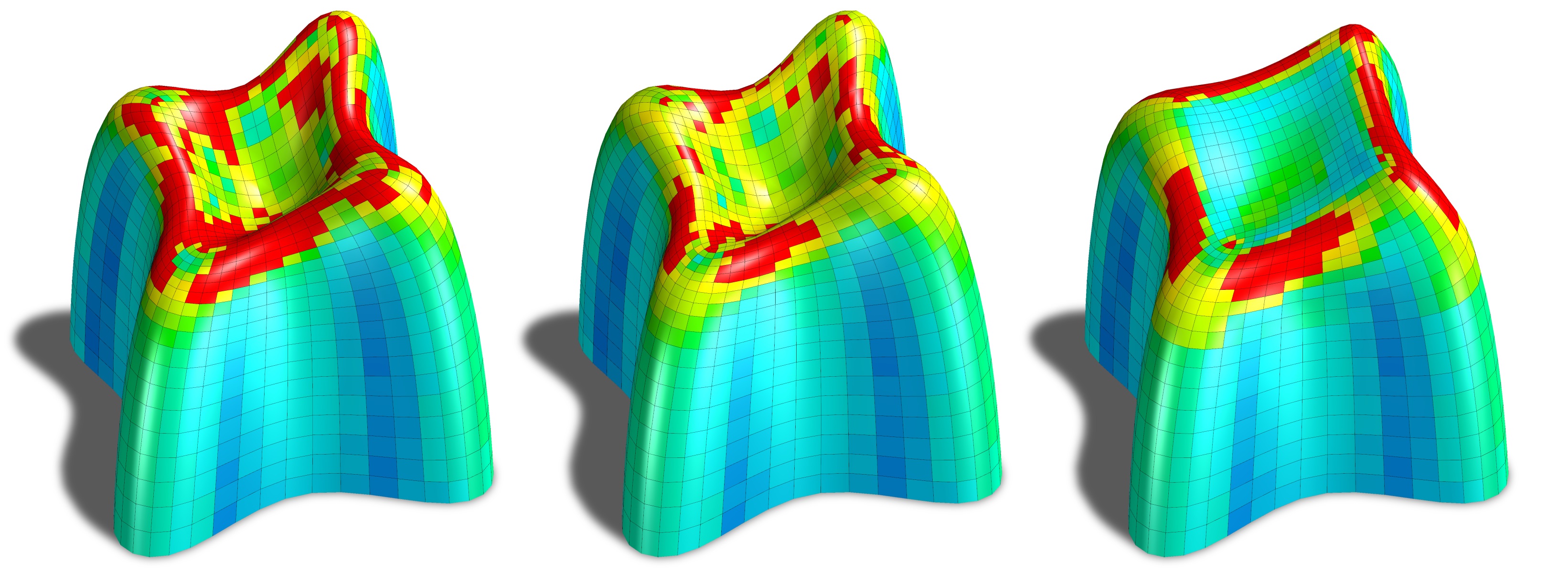}
    \put(80,-2){\stressbar}
    \put(6,32){a}
    \put(38,32){b}
    \put(70,32){c}
    \end{overpic}
    \caption{Optimization of the Lilium Tower (model by Zaha Hadid Architects) for different target properties.
    (a) Stress values for the initial panel layout.
    (b) Optimizing the design only for stress reduction and proximity to the original design leads to more panels within the stress threshold, but also to a non-smooth curve network.
    (c) Allowing the design to deviate from the input and including fairness, produces a smoother result with reduced stress.
    Number of panels exceeding $65$ MPa is, respectively, $293$, $131$, and $225$.}
    \label{fig:lilium_stress45_comp}
\end{figure}
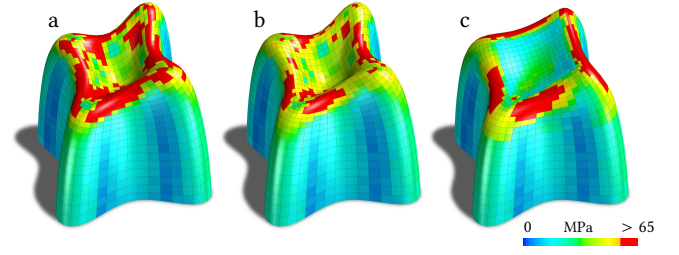

\subsubsection{Initialization}
\label{subsubsec:initialization}

The edge plane vector $\mathbf{s}_e$ of an edge $e$ is initialized so that $\Pi_e$ is the bisecting plane of two discrete osculating planes, as in the explanation for Equation~(\ref{eq:smooth}). The angles $\theta_i$ are initialized so that they are at most $5^\circ$ and so that the tangents lie as close as possible to the estimated tangent planes of the reference geometry. After initializing all other variables and computing an estimated stress value per face panel, the auxiliary variables are initialized such that they add up to the inequality constraint bound or zero otherwise (i.e., the inequality constraint is not satisfied). The shape $\mathbf{S}_{\mathbf{{p}}}$ of each panel is initialized with the MDN prediction using the initial boundary parameters $\mathbf{p}$. In case there are two possible shapes, we use the one that provides the best solution considering application-dependent criteria (e.g., stress reduction). When looking for the smoothest fit, we pick the one minimizing $\sum (1 - \mathbf{n}_i \cdot \mathbf{n}_e)^2$, i.e., a measure of angle deviation between each edge normal (sum of two adjacent face normals orthonormalized to $\mathbf{e}$) and the surface normal $\mathbf{n}_i$ evaluated at the parameter $t=0.5$ of the edge curve.
 
\subsubsection{Optimization weights}
 
The weights associated with the target functional $\mathcal{E}$ act as handles for the designer to guide the output of the optimization toward the desired result. We do not opt for a fixed weight configuration because the ideal balance is not uniquely defined, but is instead governed by project-dependent factors such as budget and design ambition. 
 
In all our experiments, we found it sufficient to assign the weights either to zero or to the values $10^{\{-2, -1, 0\}}$. Figure~\ref{fig:lilium_stress45_comp} shows one example of the different effects possible when changing the property importance. In practice, and as a rule of thumb for a standard optimization where we prioritize stress reduction and smooth panels (in that order), we use weight values $w_\sigma = 1$, $w_\text{s} = w_\text{c} = 10^{-2}$, and $w_\text{p} = w_{f} = 10^{-1}$. We also reduce the fairness importance at the $i$-th optimization iteration by scaling its weight by $0.9^i$.

%-------------------------------------------------------------------
\begin{figure}[b!]
   \includegraphics[width=\linewidth]{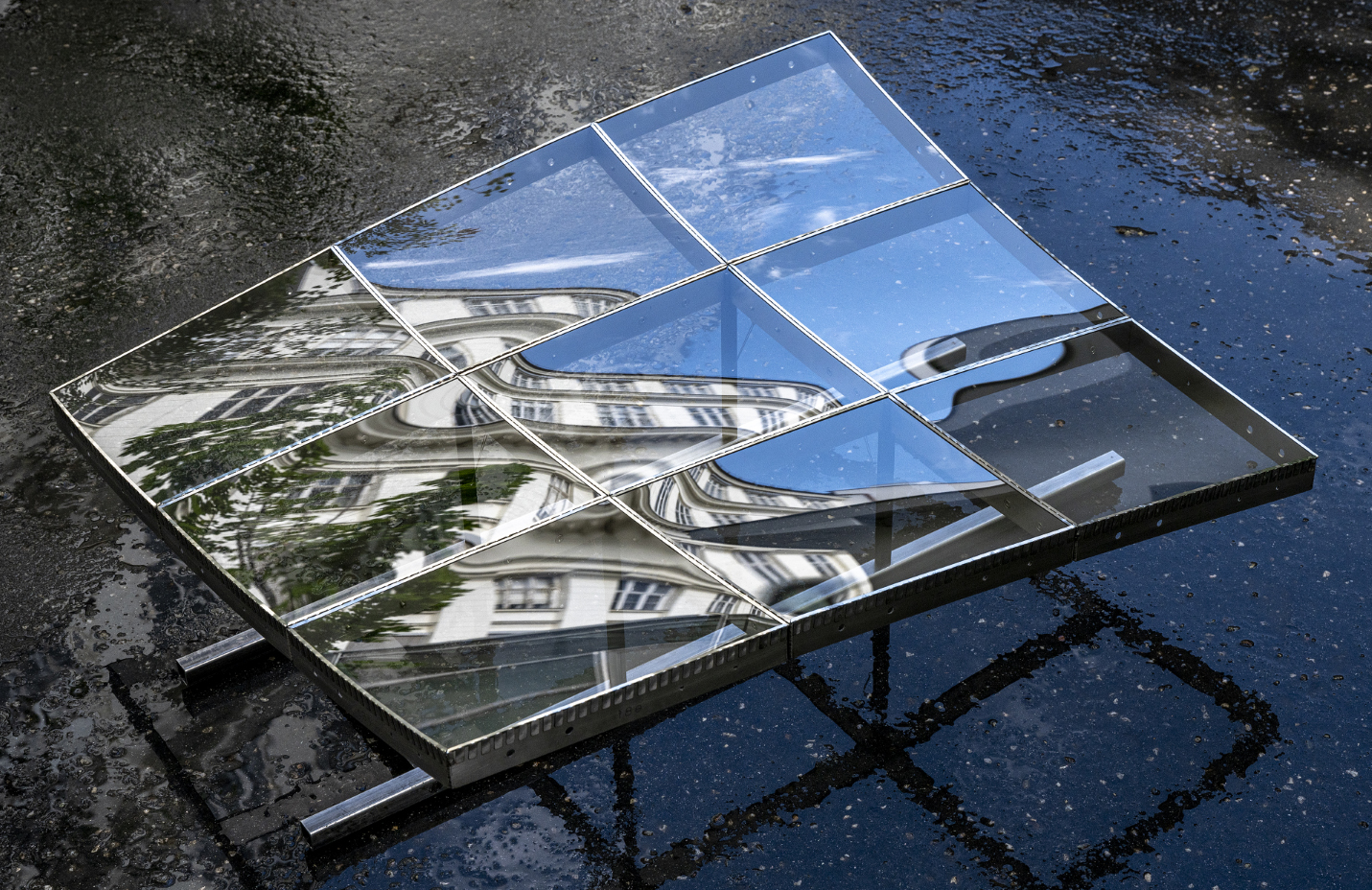}
   \caption{Realization of a doubly curved surface using 3x3 cold bent panels.}
    \label{fig:fabricated_surface}
\end{figure}

%-------------------------------------------------------------------

\begin{figure*}[ht]
    \centering
    \includegraphics[width=0.25\linewidth]{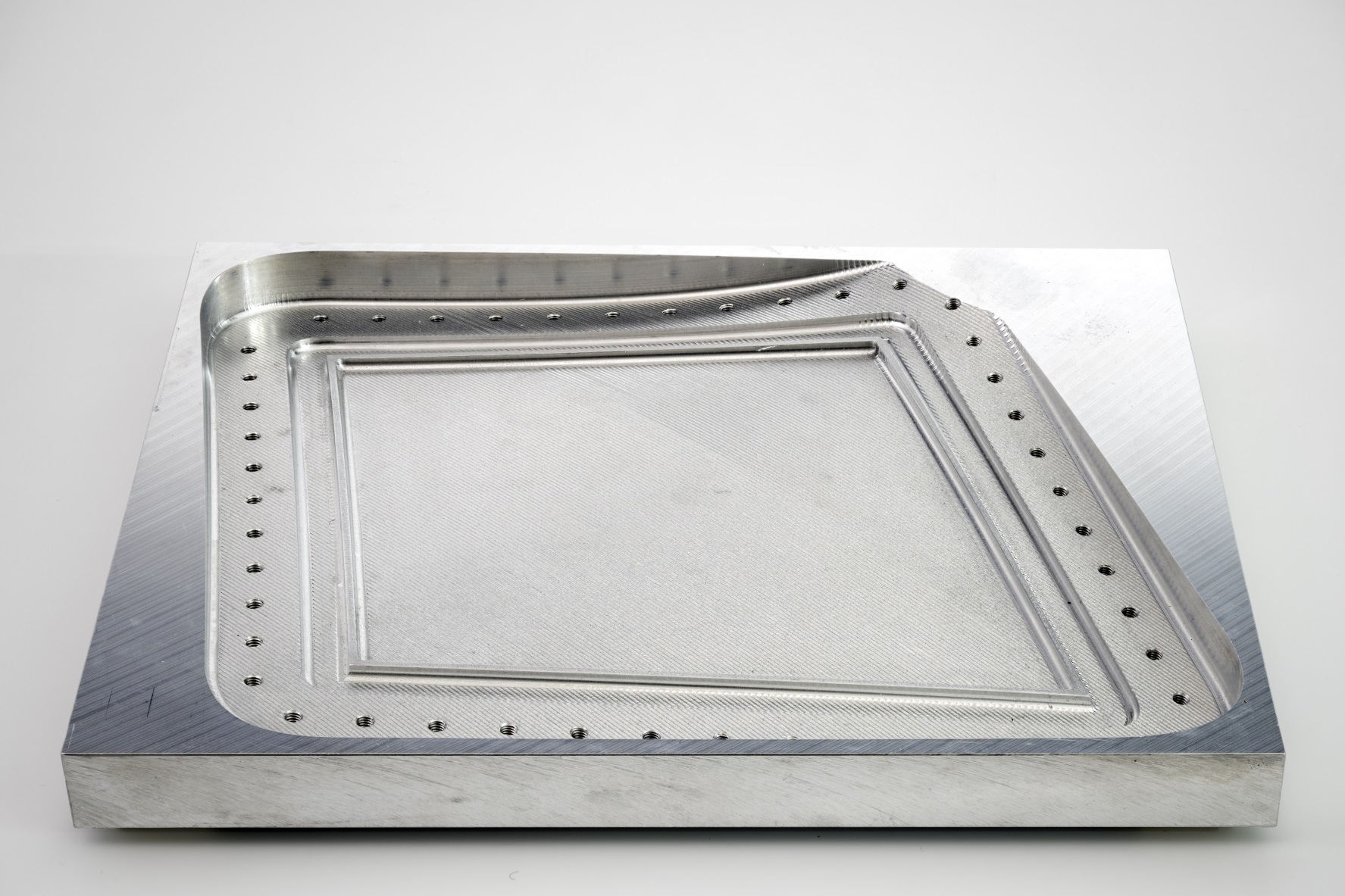}
    \includegraphics[width=0.25\linewidth]{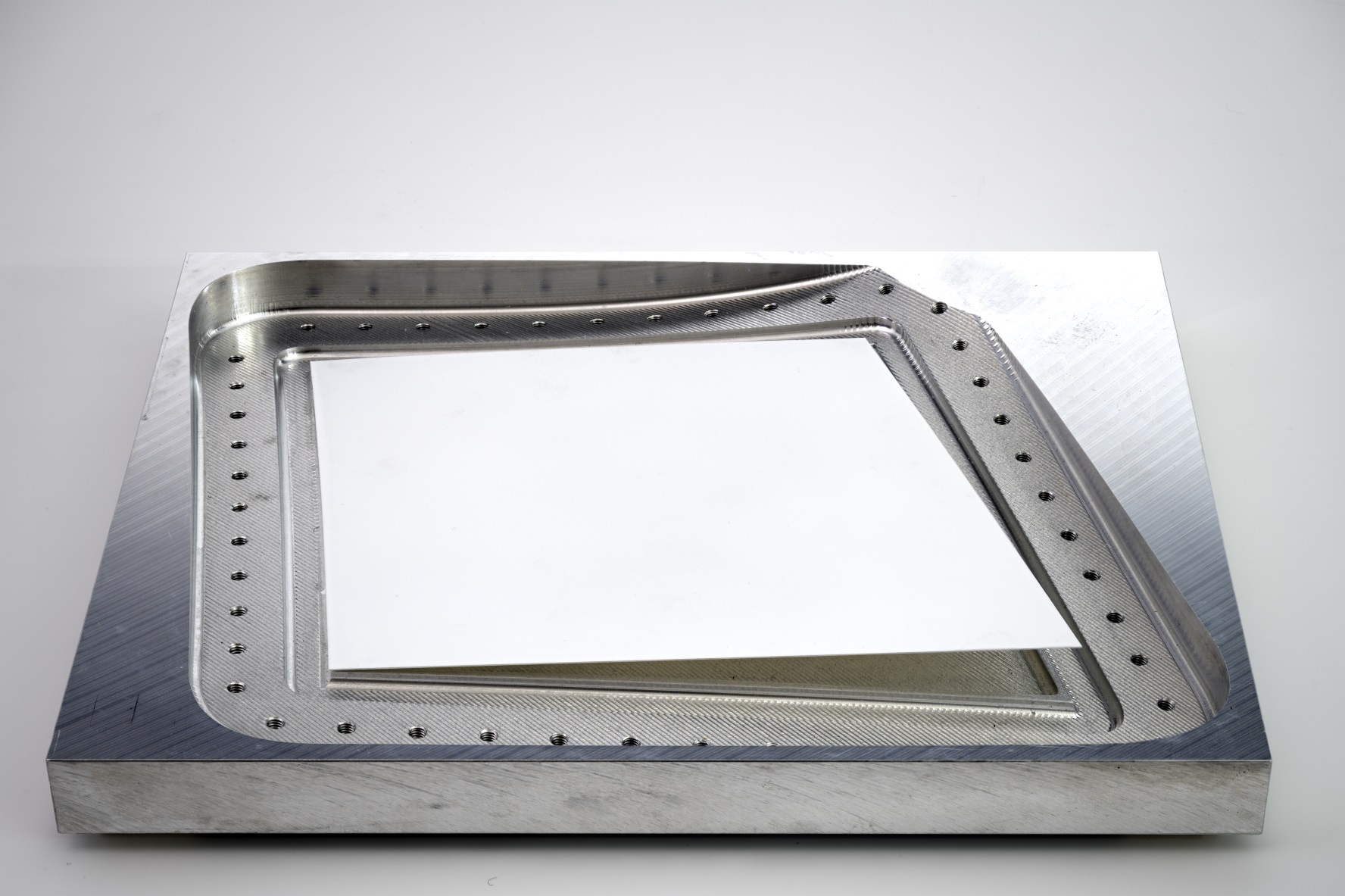}
    \includegraphics[width=0.25\linewidth]{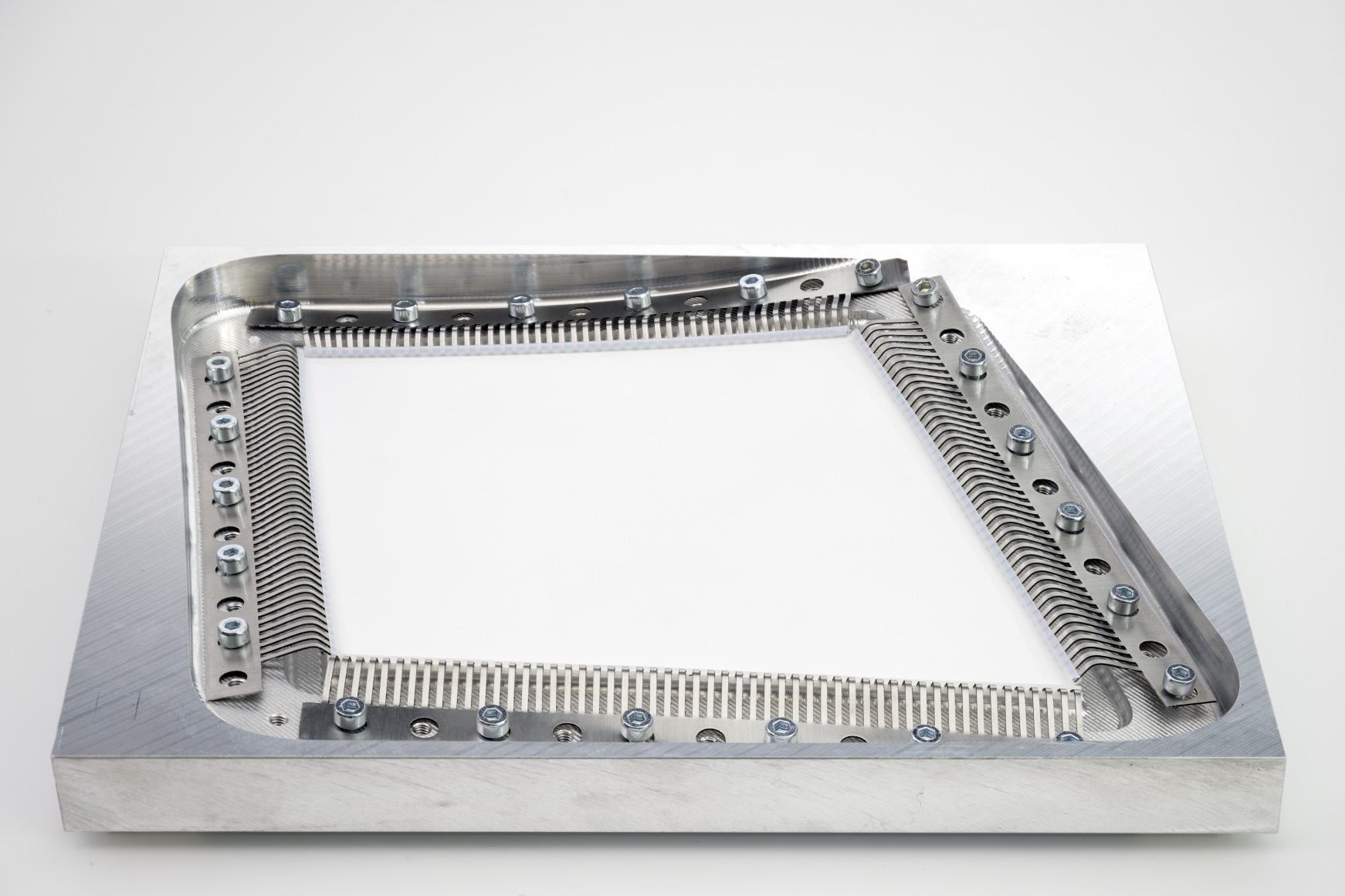}
    \begin{overpic}[width=0.23\linewidth]{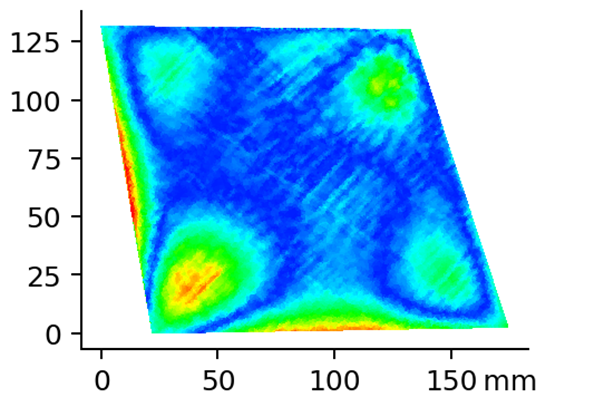}
    \put(60,66){\offsetbar}
    \end{overpic}
    \caption{A doubly curved panel of a thickness of 0.35~mm with off-plane corner deviation of 6.9~mm. A white coating has been applied to it for 3D-scanning. Right: deviation from the simulation by at most 0.12 mm.}
    \label{fig:assembly_process}
\end{figure*}

\section{Results}

\subsection{Experimental validation}

We experimentally validated our simulation results and design workflow. For practical reasons, the experiments were done at a small scale using borosilicate thin glass of about $180\times130$~mm$^2$ and 0.35~mm thickness.

For the validation of the simulation results (see Figure~\ref{fig:assembly_process}), high precision frames were machined from cast aluminum. The glass panel is pressed down on a 2~mm wide smooth support frame by a dense array of stainless steel finger springs which are cushioned by 0.5~mm polytetrafluoroethylene (PTFE). The support frame matches a thin boundary strip of the simulated glass panel. To test the accuracy of the predicted shape, we selected and 3D-scanned a panel with a high estimated maximal stress (98 MPa), which is beyond our safety limit but still manufacturable (see Figure~\ref{fig:assembly_process}). The obtained surface was registered to the output of our shape optimization routine, and we observed a worst-case deviation of 0.12~mm. Note that we registered an offset surface from the optimal mid-surface to account for the glass thickness.

The frames for the design model created with our tool are illustrated in Figure~\ref{fig:fabricated_surface} and were built from laser cut and welded 1.2~mm thick stainless steel sheet metal. The glass, which is cushioned by tape, is pressed down on to the frame by L-shaped stainless steel fixtures spot welded to the frame. The presented design model is negatively curved and consists of nine individual panels, each about $200\times170$~mm$^2$ in size. The expected stress levels range from 20 to 62~MPa. As predicted, all panels are fabricable and intact.

During bending, our panels usually do not need to go through more extreme deformations than the final one, meaning that we do not expect higher stresses while bending. Because normally panels have a dominant bending direction, we observed that it is possible to ``roll'' the glass onto the frame accordingly during assembly. Clamping the glass to the frame fixes the normals at the boundary, which makes one of the alternative shapes preferable.

%-------------------------------------------------------------------
\begin{figure}[b]
    \begin{overpic}[width=0.85\linewidth]{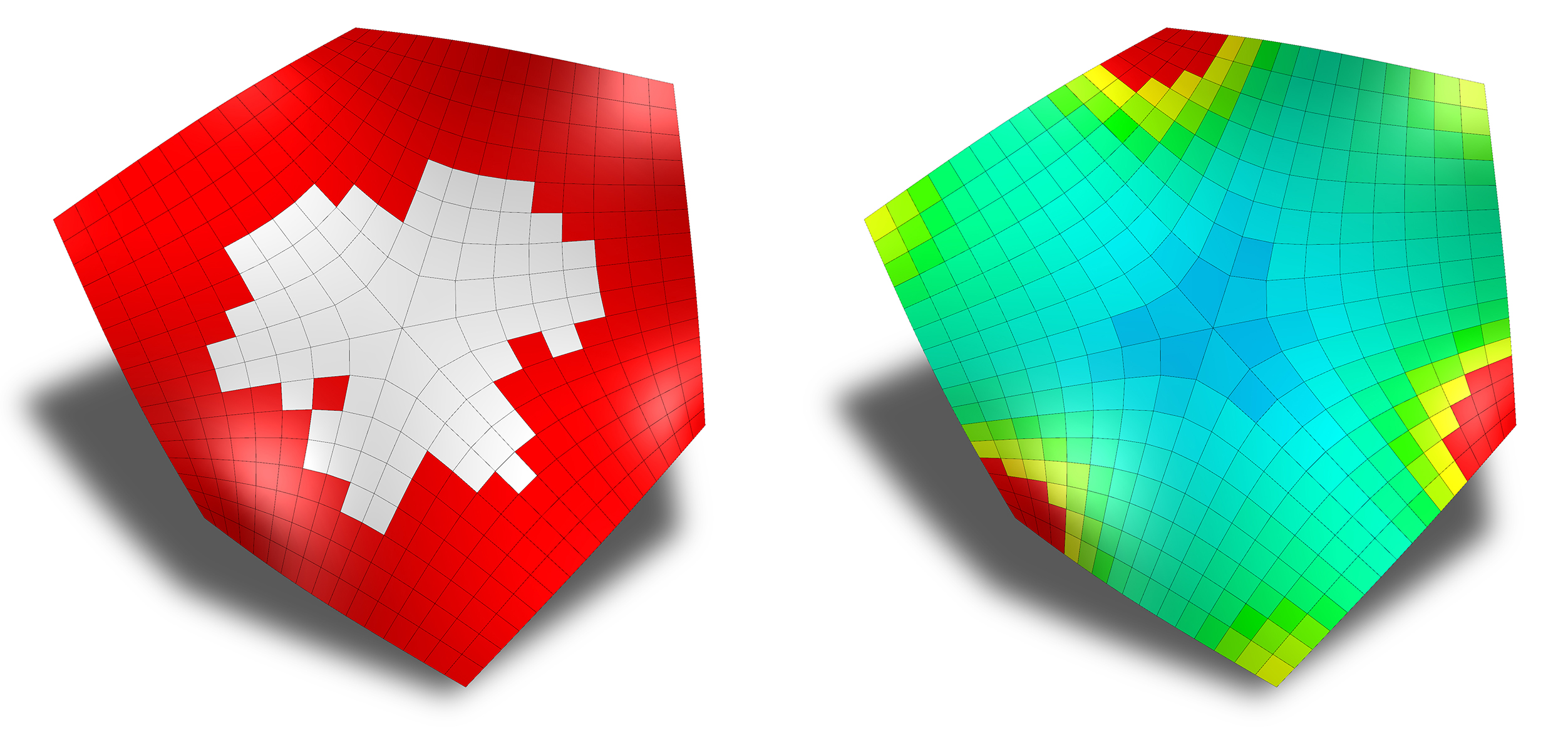}
    \put(52,-2.5){\stressbar}
    \put(3,37){a}
    \put(55,37){b}
    \end{overpic}
   \caption{Bent glass capabilities.
    (a) A quadrilateral mesh where the red faces exceed a deviation of a planarity of $0.02$ (measured as the distance between the diagonals divided by average edge length) and, therefore, not suitable for a flat glass panelization.
    (b) A cold bent panelization with corresponding face stresses.
    The stress values for the six central panels have been computed via simulation because they were outside the MDN input domain.
    According to a stress limit of $65$ MPa, most of the panels optimized are feasible.
    The resulting cold bent panelization is shown in Figure~\ref{fig:render_comb}.}
    \label{fig:monkey_saddle}
\end{figure}

\subsection{Validation of data-driven model}

Our data-driven model (Section~\ref{sec:learning}) must reproduce the output of the physical simulation model efficiently and accurately. To evaluate its accuracy, we generate a test set of 10K panel boundaries and use the data-driven model to predict the conforming surfaces. We consider only admissible surfaces, i.e. those with a predicted probability of at least 5\%. The surface predictions are used to initialize our physical shape optimization routine to obtain the true shape and stress values for a comparison. In the resulting test set, the mean maximal stress value is 83 MPa, and the standard deviation is 52 MPa with 57\% of panels having a maximal stress value above the threshold.

We evaluate the shape prediction on panels with maximal stress below 65~MPa, which results in a mean average error (MAE) of $\sim$0.5~mm. Note that this is significantly less than the assumed 1~mm thickness of the glass. We evaluate stress on panels whose true maximal stress is in the range 50--65~MPa (our region of interest); our predictions have MAE of $\sim$2.9~MPa. Moreover, the 67th percentile error is 2.5 MPa, and the 88th percentile error is 5 MPa. In addition, we evaluate how often our model correctly predicts whether or not the actual maximal stress value (in contrast to the $L_p$-norm) exceeds the 65 MPa threshold. From our test set, we obtained below 1\% of false negatives (when the model incorrectly predicts the panel is feasible) and 15\% of false positives.

%-------------------------------------------------------------------
\begin{figure*}[t]
    \centering
    \includegraphics[width=1\linewidth]{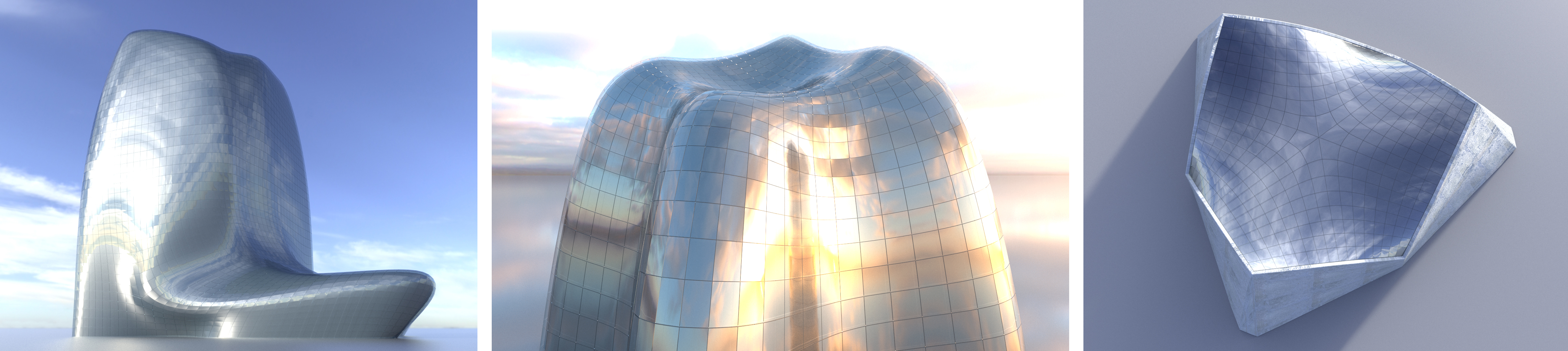}
    \caption{Dominant cold bent glass realizations of the NHHQ model (left). The Lilium Tower (center) after optimization for
    smoothness and stress reduction. The surface from Figure~\ref{fig:monkey_saddle} as an architectural design (right). Panels exceeding the maximum stress (check Figures \ref{fig:nhhq_stress}, \ref{fig:lilium_stress45_comp}, \ref{fig:monkey_saddle}) are realized with hot bending.}
    \label{fig:render_comb}
\end{figure*}
%-------------------------------------------------------------------

\subsection{Applications}

From a manufacturing point of view, the simplest solution to clad architectural surfaces is the use of planar panels. However, this simplification sacrifices the visual smoothness of the surface. Moreover, planar panels impose a restriction on the panels layout, and in negatively curved areas, there is often no other choice than to follow the principal curvature directions of the surface. On the other hand, a panelization with doubly curved panels is often prohibitive because of the high production cost of custom molds. Cold bent glass can be then a suitable solution. In Figure~\ref{fig:curtain_full_comp}, we compare the visual appearance of the smoothest possible panelization achievable with planar panels with a cold bent one, while in Figure~\ref{fig:monkey_saddle}, we show a panelization layout that is mostly feasible with cold bent glass, but not with planar panels. In the following, we illustrate how users can employ our workflow for architectural panelization and design.

\subsubsection{Façade panelization}\label{sub:façade_panelization}

In this case, the input is a quadrilateral mesh that encodes both the design shape and the panel layout. Once the edges of the input mesh are smoothed via cubic B\'ezier curves, we can predict the panels' shapes and stresses. Those panels exceeding the failure criterion shall be realized with custom molds. At this point, the user can optimize the shape for the reduction of stress and kinks between panels, and tune the weights described in Section~\ref{ssub:numerical_optimization_algorithm} for choosing an appropriate compromise between fidelity with the original shape, number of custom molds needed, and visual smoothness (see Figures~\ref{fig:nhhq_stress}, \ref{fig:nhhq_kink}, and \ref{fig:lilium_stress45_comp}). To show cold bent glass' capabilities in façade panelization, we tested this workflow on the challenging NHHQ and Lilium Tower models by Zaha Hadid Architects, which were never realized. The initial quadrangulations for these two designs are planar quad meshes which were created by the architect. The results are shown in Figure~\ref{fig:render_comb}.
 
\subsubsection{Façade design}

Besides the panelization of a given shape, our workflow is very well suited as an interactive design tool. In this case, the user can interactively modify the quad mesh that represents the panel layout and gets immediate feedback on which panels can be produced with cold bent glass, while exploring different designs. Initial values are computed as in \ref{subsubsec:initialization} and we use the mesh vertex normals for the estimated tangent planes at the vertices. The estimation times are compatible with an interactive design session. Once the user is satisfied with a first approximate result, the panelization can be further optimized, as described in Section~\ref{sub:façade_panelization}, to improve the smoothness and reduce panel stresses. In this step, we can further reduce the number of panels that are not feasible for cold bending. Figures~\ref{fig:teaser}, \ref{fig:ring}, and \ref{fig:monkey_saddle} show some sample architectures designed with this procedure. Furthermore, the accompanying video demonstrates the interactive feedback capabilities of our system for efficient form finding and exploration of the constrained design space while keeping the designer in the loop.

All interactive design sessions were performed on an Intel\textregistered{}  Core\texttrademark{} i7-6700HQ CPU at $2.60$ GHz and NVIDIA GeForce GTX~960M. The MDN is implemented in TensorFlow~2.1 and is run on the GPU. For 1K panels, the prediction time is $0.1$ seconds while the optimization averages $3$ seconds per iteration. We usually deal with less panels because we target the selected high-stress areas of the overall design. A total of 10--20 iterations are enough for the desired results. In comparison, our shape optimization, as described in Section~\ref{sec:simulation}, implemented in C++ and using the IPopt optimization library with code-generated derivatives takes around 35~seconds on average for a single panel with $\sim10^3$ elements. Note that this routine is not fully optimized for speed because it is not required during the interactive phase but mainly used for acquiring training data.

\section{Discussion and Conclusion}

We have introduced an interactive, data-driven approach for material-aware form finding of cold bent glass façades. It can be seamlessly integrated into a typical architectural design pipeline, allows non-expert users to interactively edit a parametric surface while providing real-time feedback on the deformed shape and maximum stress of cold bent glass panels, and it can automatically optimize façades for fairness criteria while maximal stresses are kept within glass limits. Our method is based on a deep neural network architecture and multi-modal regression model. By coupling geometric design and fabrication-aware design, we believe our system will provide a novel and practical workflow, allowing to efficiently find a compromise between economic, aesthetic, and engineering aspects.

Identifying such a compromise usually involves multiple competing design goals. Although we have demonstrated the applicability of our system for several design criteria, it would be interesting to extend the design workflow by adding capabilities, for example, for strictly local edits, marking some panels as a priori hot bent or specifying kink edges. Because of our differentiable network architecture, in theory, it should be trivial to incorporate additional criteria into our optimization target functional or even employ a different numerical optimization algorithm if desired. 

Similar to all data-driven techniques, we should only expect accurate predictions from our network if similar training data was available. Surprisingly, we noted that we were able to discover stable states that we initially did not find with the traditional optimization approach, and used these to enrich our database. Each boundary in the training set is associated with a single stable surface output by the simulator. Nevertheless, the network may correctly predict the existence of a second stable state for that boundary, because its predictions implicitly incorporate information from similar panels in the training set, where the second state is seen. However, we cannot guarantee that our database contains all relevant stable states and that all of them will be predicted. Identifying {\it all} stable states and optimally sampling the database using this information would be an interesting avenue for future work. For fabrication, in our experiments, reproducing the desired particular state was trivial and emerged when intuitively attaching the glass to the frame. 

In the presence of more than one potential state, our system currently selects in each iteration per panel the state that best fits our application-dependent criteria. An alternative would be to compute a global, combinatorial optimal solution among all potential states. However, because of the combinatorial complexity, this would result in a much harder and probably computationally intractable optimization problem. We also considered solving the combinatorial problem by using a continuous relaxation but ultimately did not find evidence in our experiments that would indicate the need for such an approach as we observed stable convergence to satisfactory results. However, identifying the global minimum would nevertheless be an interesting research challenge. 

By design, our workflow is not limited to a particular shell model, and in theory, more advanced and extensively experimentally validated engineering models could be used if needed. We believe the benefit of our learning-based approach would even be more evident with more complex mechanical models, because they are computationally significantly more expensive. Our workflow could serve as an inspiration for many other material-aware design problems. For future work, it would be exciting to explore extensions to different materials, for instance metal, wood, or programmable matter that can respond to external stimuli, such as shape memory polymers or thermo-reactive materials.

\begin{acks}
	We thank IST Austria's Scientific Computing team for their support, Corinna Datsiou and Sophie Pennetier for their expert input on the practical applications of cold bent glass, and  Zaha Hadid Architects and Waagner Biro for providing the architectural datasets. \href{https://www.flickr.com/photos/fran001/49628199916/}{Photo} of Fondation Louis Vuitton by Francisco Anzola / CC BY 2.0 / cropped. Photo of Opus by Danica O. Kus.
	
	This project has received funding from the \grantsponsor{H2020-MSCA-ITN-2015}{European Union's Horizon 2020 research and innovation program}{https://ec.europa.eu/programmes/horizon2020/en} under grant agreement No~\grantnum[]{H2020-MSCA-ITN-2015}{675789} - Algebraic Representations in Computer-Aided Design for complEx Shapes (ARCADES), from the \grantsponsor{ERC-2016-STG}{European Research Council}{https://erc.europa.eu/} (ERC) under grant agreement No~\grantnum{ERC-2016-STG}{715767} - MATERIALIZABLE: Intelligent fabrication-oriented Computational Design and Modeling, and SFB-Transregio ``Discretization in Geometry and Dynamics'' through grant \grantnum{FWF}{I 2978} of the \grantsponsor{FWF}{Austrian Science Fund}{https://fwf.ac.at/en/} (FWF). F. Rist and K. Gavriil have been partially supported by \grantsponsor{KAUST}{KAUST}{https://kaust.edu.sa/en} baseline funding.
\end{acks}

%%%%% BODY %%%%%

%%%%% REFS %%%%%

\bibliographystyle{ACM-Reference-Format}
\bibliography{glassbending}

%%%%% REFS %%%%%

\appendix

\section{Sampling panel boundaries}
\label{app:boundary-sampling}

We briefly describe how the panel boundaries forming the training set for our data-driven model are sampled (Section~\ref{sec:ddm-initial-dataset}). We parameterize panel boundaries invariantly to rigid transformations, by corner pairwise squared distances $\mathbf{d}$, edge-plane inclinations $\bm\gamma$, and halfedge tangent directions $\bm\theta$ (Section~\ref{sec:geometry}). In order to sample $\mathbf{d}$ such that it represents a valid quad, we start with two adjacent edge lengths $l_1$, $l_2$, an angle $\alpha$ between them, and a displacement $\mathbf{a}$ of the remaining vertex from the point that would form a parallelogram. We sample each of these parameters as follows:
\begin{itemize}
\item $l_1, \, l_2 \sim \mathrm{Uniform}[0.15,0.60]$; this corresponds to 15--60~cm for a 1~mm thick panel,
\item $\alpha \sim \mathrm{Uniform}[60^\circ,120^\circ]$,
\item $\mathbf{a}$ is given by sampling a point on the unit sphere, then scaling it by a factor drawn from $\mathrm{Uniform}[0,\min\{l_1,l_2\}/4]$,
\item $\gamma_i \sim \mathrm{Uniform}[-90^\circ,90^\circ]$,
\item $\theta_i$ is given by $\arccos$ of a value sampled from $\mathrm{Uniform}[\cos 5^\circ, 1]$, negated with probability $1/2$, so $\theta_i \in [-5^\circ,5^\circ]$.
\end{itemize}

Note that our model for the deformed shape and stress is invariant under scaling of all geometric magnitudes.
Our sampling ranges are chosen to allow scaling the results to thickness/curvature ratios commonly used in cold bent glass façades.

\end{document}